\documentclass[10pt]{article}
\usepackage[margin=1in]{geometry}
\title{Measuring Justice in Machine Learning}
\author{Alan Lundgard\\
Massachusetts Institute of Technology\\
Cambridge, MA, United States\\
\href{mailto:lundgard@mit.edu}{lundgard@mit.edu}
}
\date{30 September 2020\footnote{
    Draft version 0.3.5. This paper was presented at the ACM Conference on Fairness, Accountability, and Transparency (30 January 2020) and at the ACM SIGACCESS Conference on Computers and Accessibility: Workshop on AI Fairness for People with Disabilities (27 October 2019). \copyright 2020 Alan Lundgard. All rights reserved. Please cite with the following link: \url{https://dl.acm.org/doi/10.1145/3351095.3372838}. 
}}

\usepackage[round]{natbib}
\usepackage[dvipsnames,table]{xcolor}
\usepackage[linktoc=all,colorlinks=true,linkcolor=Mahogany, urlcolor=magenta, citecolor=ForestGreen]{hyperref} 
\usepackage{breakurl}
\usepackage[capitalise, nameinlink]{cleveref}
\usepackage[nottoc,numbib]{tocbibind} 
\usepackage[nolist,nohyperlinks]{acronym}

\begin{acronym}
    \acro{fair ML}{fair machine learning}
\end{acronym}
\usepackage{textcomp}

\usepackage{hhline}
\newcolumntype{?}{!{\vrule width 2pt}}

\providecommand{\keywords}[1]{\noindent\textbf{\textit{Keywords}} #1}

\providecommand{\sectionlabel}[1]{(\cref{#1}: \nameref{#1})}

\begin{document}
\maketitle
\begin{abstract}\noindent
How can we build more just machine learning systems? To answer this question, we need to know both what justice is and how to tell whether one system is more or less just than another. That is, we need both a definition and a measure of justice. Theories of distributive justice hold that justice can be measured (in part) in terms of the fair distribution of benefits and burdens across people in society. Recently, the field known as fair machine learning has turned to John Rawls's theory of distributive justice for inspiration and operationalization. However, philosophers known as capability theorists have long argued that Rawls's theory uses the wrong measure of justice, thereby encoding biases against people with disabilities. If these theorists are right, is it possible to operationalize Rawls's theory in machine learning systems without also encoding its biases? In this paper, I draw on examples from fair machine learning to suggest that the answer to this question is no: the capability theorists' arguments against Rawls's theory carry over into machine learning systems. But capability theorists don't only argue that Rawls's theory uses the wrong measure, they also offer an alternative measure. Which measure of justice is right? And has fair machine learning been using the wrong one?
\end{abstract}

\newpage

\tableofcontents
\newpage 

\section{Introduction}
\label{section:1}

Theories of distributive justice hold that justice can be measured (in part) in terms of the fair distribution of benefits and burdens across people in society.
These theories help us (i.e., community members, researchers, and policymakers) to know both what justice is and how to tell whether one society or system is more or less just than another.
Aiming to mitigate ``algorithmic unfairness'' or bias, the field known as \ac{fair ML} has recently sought to operationalize theories of distributive justice in ML systems~\citep{binns_fairness_2018, dwork_fairness_2011}.
Researchers have quantitatively formalized theories of equal opportunity~\citep{hardt_equality_2016, heidari_fairness_2018, joseph_fairness_2016}, as well as portions of John Rawls's influential theory of distributive justice~\citep{hashimoto_fairness_2018, jabbari_fairness_2017, joseph_rawlsian_2016}.

While the quantitative formalization of these theories has been on-going over the past 30-plus years, in theory we don't need to assume a quantitative measure at the outset~\citep{roemer_theories_1998}. Historically, formalization has been accompanied by robust philosophical debate over the question ``\emph{what} is the right measure of justice?'' That is, when evaluating whether a system is fair or just, what type of benefit or burden should we look at, as a measure? And how should we approach measuring that benefit, quantitatively or otherwise? These questions are important because different benefits affect the well-being of different people in different ways, and different theories offer different approaches to measurement. Selecting the wrong measure, or operationalizing the wrong theory, may prematurely foreclose consideration of alternatives, disregard the well-being of certain people, or reproduce and entrench---rather than remediate---existing social injustices.

Case in point, \ac{fair ML} researchers have often conceptualized algorithmic unfairness as a problem of ``resource allocation'' wherein they aim to mitigate unfairness by algorithmically reallocating resources across people~\citep{donahue_fairness_2020,hashimoto_fairness_2018}. For inspiration and operationalization, some well-intentioned researchers have turned to the theories of distributive justice that similarly aim to remediate social injustices by quantifying, measuring, and reallocating resources across people in society~\citep{hoffmann_where_2019}. Notably, the theories of John~\cite{rawls_theory_1971}---recently called ``artificial intelligence's favorite philosopher'' due his popularity in \ac{fair ML}~\citep{procaccia_ai_2019}---have received significant attention.

However, when unfairness or injustice is conceptualized as a problem of resource allocation, the answer to the question ``\emph{what} is the right measure?'' is already embedded in the statement of the problem. That is, ``resource allocation'' problems entail a particular solution: to mitigate unfairness or remediate social injustice, it is \emph{resources} that are quantified, measured, and reallocated across people. But are resources the right type of benefit to look at, as a measure of justice? Rawls answers yes, with important qualifications. Other philosophers known as capability theorists answer no. They argue that resource measures encode biases against disabled and other socially marginalized people. Rawls mounts a defense of resources, but capability theorists don't only argue against resources, they also offer an alternative measure: \emph{capabilities}. Which measure of justice is right? And has \ac{fair ML} been using the wrong one?

\subsection{Overview of Contents}

In this paper, I extend---from philosophy to machine learning---a longstanding debate between defenders of two prominent approaches to measuring justice: the \textbf{resourcist approach} (defended by philosophers like John Rawls, Ronald Dworkin, and Thomas Pogge) and the \textbf{capability approach} (defended by philosophers like Amartya Sen, Martha Nussbaum, and Elizabeth Anderson). I spell out the differences between these competing approaches, and how they have been (or could be) operationalized in the design and development of ML systems. First, I inquire into what makes the resourcist approach amenable to operationalization in these systems. Then, I inquire into whether the capability approach could be operationalized with more success. The inquiry proceeds by way of the capability theorists' critiques of the resourcist approach. 

At a high level, the inquiry proceeds as follows. In the resources-versus-capabilities debate, the two approaches are distinguished based on certain criteria for a measure of justice. I begin by introducing these criteria, as well as the philosophical tools and terminology of distributive justice~\sectionlabel{section:2}. Based on these criteria, and their underlying characteristics, I establish a theoretical resemblance between resource measures and the measures often adopted in \ac{fair ML}, where algorithmic unfairness has often been conceptualized as a problem of resource allocation. This resemblance suggests how critiques made by the capability theorists might carry over to the ML systems that operationalize Rawls's theory~\sectionlabel{section:3}. To show how these critiques do carry over, I discuss two real-world case studies from ML. In these cases, I argue that the measures selected for mitigating algorithmic unfairness end up entrenching and reproducing social injustices---just as the capability theorists argue resource measures do~\sectionlabel{section:4}. As an alternative to operationalizing the resourcist approach in ML systems, I introduce the capability approach~\sectionlabel{section:5}. I outline what operationalizing this approach could look like in practice---through a participatory and democratic process---and discuss how doing so could better remediate the social injustices from the case studies, and beyond. I also discuss some of the most noteworthy advantages and limitations of the capability approach~\sectionlabel{section:6}. Finally, I conclude by replying to some potential objections~\sectionlabel{section:7}.

Extending the longstanding resources-versus-capabilities debate from philosophy to ML reveals some broadly applicable characteristics of (and criteria for) a measure of justice in machine learning.
\begin{itemize}
    \setlength\itemsep{0.0em}
    \item Whether a measure is single or multi-valued; quantitative and/or qualitative.
    \item Whether what is being measured is a ``means to an end'' or an ``end in itself.''
    \item Whether or not a measure is sensitive to people's heterogeneities and societal contexts.
    \item Whether or not a measure is publicly legible and verifiable.
\end{itemize}
\noindent
These characteristics and criteria have broad implications for the practicability of quantitative and/or qualitative evaluations of justice and fairness in machine learning, for the participatory design of these systems, and for the democratic oversight over the institutions that build and deploy them.
\section{Theories of Distributive Justice}
\label{section:2}
There are many theories of justice---distributive, legal, restorative, retributive, among others---all uniquely important and often complementary. Here, I focus on theories of distributive justice. While some of these theories (namely, Rawls's resourcist approach) have recently been operationalized in ML systems, prominent alternatives (such as the capability approach) seem to have been sidelined.
To understand the differences between these two approaches, and how they have been (or could be) operationalized in \ac{fair ML}, we first need to know more about how theories of distributive justice work.

Theories of distributive justice propose \emph{principles} for fairly distributing benefits (e.g., material resources, services, and opportunities) and burdens (e.g., barriers, taxes, and punishments) across people in society. These principles provide normative guidance for a society's systems and institutions to follow, to detect and remediate historical and on-going social injustices. Distributive principles are specified by two features: their measure and their rule. The \textbf{measure} specifies \emph{what} type of benefit or burden is to be distributed. The \textbf{rule} specifies \emph{how} that benefit or burden is to be distributed~\citep{brighouse_justifying_2010}. For example, as I noted in the Introduction, to remediate social injustices, some philosophers propose reallocating resources across people in society. Similarly, to mitigate algorithmic unfairness, some \ac{fair ML} researchers propose the same. But the principle of ``reallocate resources'' is specified both too quickly and not yet fully. It is too quick because we don't need to select resources as the type of benefit to be reallocated (the measure). It is not yet fully-specified because it also matters how resources are to be reallocated (the rule). We can specify both the measure and the rule in many ways. And each feature combination will have different normative implications, lead to different distributions, and affect the well-being of different people differently.\footnote{
 Unlike distributive principles from philosophy, the field of \ac{fair ML} has not made a consistent conceptual and terminological distinction between the measure (\emph{what} is being distributed) and the rule (\emph{how} it is being distributed). Instead, these two features have often been collapsed into a single conception, referred to as the ``fairness metric''. For example, ``equal predictive accuracy'' is one such fairness metric, and there are countless others on offer~\citep{narayanan_21_2018,deborah_hellman_measuring_2019}. To be consistent with the philosophical terminology, I will refer to equal predictive accuracy (and the like), not as a fairness metric, but as a distributive principle that specifies predictive accuracy as its measure and equality as its rule. Making this conceptual and terminological distinction is important because it helps us to be clearer about which feature (the measure or the rule) contributes to the unfairness or unjustness of a distribution. Failure to make this distinction can lead us to take either feature for granted.
}

Theories of distributive justice primarily differ on which measure-rule combination they defend as the most successful remediator of social injustice. Some philosophers argue that resources like monetary income and wealth (the measure) should be distributed more-or-less equally (the rule) across people in society. Others defend different measures or rules, arguing that the principle of ``equal income and wealth'' is not an adequate remediator of social injustice. They argue this, not necessarily because equalizing income and wealth wouldn't be helpful at all, but because successful remediation could require that we look at resources other than money, or it could require that we look at benefits other than resources, as a measure of justice.\footnote{
 Apart from looking at the distribution of resources like income and wealth, successful remediation of social injustices might require that people in society have a high quality of life overall,  that they are paid what they deserve for their hard work, that they are not disadvantaged for being born into historically marginalized communities, that they are not subject to coercion or fraud, or that they are capable of functioning as equal citizens in a democratic society. Equalizing monetary resources like income and wealth may not remediate injustices arising from these other non-monetary, less-readily-quantifiable types of benefit and burden, so philosophers of distributive justice have defended many non-monetary, non-resource measures (such as utility, desert, luck, entitlements, capabilities, among others) on the grounds that they will do a better job.
}

Theories of distributive justice also differ on whether they consider the \emph{practicability} of a distributive principle in society. That is, remediating social injustice via redistribution doesn't only depend on specifying the right measure and rule. It also depends on whether a distributive principle can be put into practice, and under which social conditions. For some philosophers, a principle only needs to work ``in theory,'' under abstract and ideal social conditions. For others, a principle only needs to work ``in practice,'' under existing social conditions, here and now~\citep{schmidtz_nonideal_2011}. Whether a principle works in theory, in practice, or in both, is not always clear. But, we (i.e., community members, researchers, and policymakers) should at least be clear about the conditions under which a principle is \emph{intended} to work. Putting into practice a principle that is intended for other social conditions could have unintended consequences.\footnote{
 When theorizing about justice, it is important to distinguish between ideal and non-ideal modes of theorizing. \emph{Ideal theories} (such as Rawls's) propose distributive principles for arranging societies and systems under certain ideal social conditions~\citep{wenar_john_2017}. For example, Rawls's theory assumes that the social conditions of society are cooperative, and that its members are ``normal and fully cooperating members of society over a complete life,'' among other assumptions~\citep{rawls_theory_1971}. The purpose of making these idealizing assumptions is not to ignore the difficult realities that actual people face in present-day society. Rather, the purpose is to first formulate principles workable under ideal conditions, so that we are then better equipped to formulate principles workable under non-ideal conditions, through reference to the ideal~\citep{wenar_john_2017}. At their best, ideal theories provide guidance and facilitate progress in theorizing about justice.
 By contrast, \emph{non-ideal theories} (such as the capability approach) propose actual steps that actual people can take---here and now---to address the actual problems they face, and to make present-day society a better place~\citep{schmidtz_nonideal_2011}. For these philosophers, a principle which cannot be put into practice (due to institutional, informational, or technical constraints) is not yet a serious candidate for the urgent remediation of existing social injustice~\citep{lamont_distributive_2017}. What distinguishes ideal from non-ideal theorizing is not the use of idealization \emph{per se}: non-ideal theorizing can and will idealize in various ways. Rather, what distinguishes the two is the priority given to the abstract over the actual. With non-ideal theorizing, we don't need to first formulate principles that are workable under ideal social conditions, since these principles may not be helpful or applicable under non-ideal social conditions. At their worst, ideal theorizing can distract us from, marginalize, or exclude the difficult realities that actual people face in present-day society~\citep{mills_ideal_2005}. Recent work in \ac{fair ML} has suggested moving away from ideal theorizing, toward non-ideal theorizing in the design and development of ML systems~\citep{fazelpour_algorithmic_2020}.
}

While the measure and the rule of a distributive principle are equally important, in this paper I focus on the measure, largely setting aside discussion of the rule. For a fuller discussion of both, I refer interested readers to surveys by Richard~\cite{arneson_egalitarianism_2013} and by Julian Lamont and Christi Favor~\citep{lamont_distributive_2017}. Additionally, I recommend the anthology \emph{Measuring Justice: Primary Goods and Capabilities} edited by Harry Brighouse and Ingrid Robeyns, from which I draw inspiration for this paper's title as well as much of its philosophical underpinning~\citep{brighouse_measuring_2010}.

\subsection{What Is the Right Measure of Justice?}
\label{section:2.1}

To know what measure of justice is right, we first need to know what the measure is intended to do. Essentially, the measure should encompass (as much as possible) what it is that contributes to people's well-being within their society. That is, the measure should consist of some type of benefit that people would like to have (or some burden that people would like to avoid) such that having more (or less) contributes to their well-being.\footnote{
 For readers familiar with the philosophical terminology, I clarify here that I do not use the term ``well-being'' to refer specifically to the \emph{welfarist approach} to measuring justice, which is exemplified by the theories of utilitarianism. For unfamiliar readers, I clarify that in theories of distributive justice, ``well-being'' can have various conceptions that are adopted to, as T.M.~\cite{scanlon_what_1998} puts it, ``make comparisons of how well-off people would be under various conditions, as measured by these conceptions.'' For example, the welfarist approach conceives of well-being in terms of \emph{welfare}, the resourcist approach in terms of \emph{resources}, and the capability approach in terms of \emph{capabilities}. In this paper, I largely set aside discussion of the welfarist approach, for reasons that will become clearer later on. I thank Gabriel Karger for suggesting this reference.
}

Consider again the simple monetary measure of income and wealth, a resource measure. People with more income and wealth are usually better off than people with less. Acquiring more money usually helps people to do more of the things they enjoy, to achieve greater social mobility, and to generally improve their quality of life and future prospects. As a measure of well-being, income and wealth are ubiquitous, easily understood, and frequently appear in evaluations of individual, institutional, and national well-being. So, we might intuitively think that these monetary measures get something right about measuring justice. Indeed, philosophers often refer to the measure as the ``currency'' of a distributive principle~\citep{cohen_currency_2011}. Less intuitively, however, a measure of justice doesn't need to be monetary, or even numerically quantifiable. This is because, when we look more closely at a person's societal context, we find that being monetarily wealthy is not the same as being well. A person's well-being does not depend \emph{only} on their income and wealth, but also on the physiological, psychological, social, political, and economic conditions under which they live. These heterogeneous conditions improve and diminish well-being, so a measure of justice should encompass them. But, some of these heterogeneities won't be readily expressed in monetary, quantifiable terms, so a measure should include \emph{more than} income and wealth alone. If it does not, then the measure may inadequately detect and remediate social injustices arising from heterogeneous social and political conditions.

In his Tanner Lecture, Amartya~\cite{rawls_equality_1979} argued that Rawls's measure of justice is an inadequate remediator of social injustice because it is \emph{insensitive} to people's heterogeneities and societal context. This lecture inaugurated a decades-long, still on-going debate between defenders of the resourcist and capability approaches to measuring justice. Rawls defends what is called the \emph{primary goods} measure, a measure of resources.\footnote{
    I thank Sally Haslanger for encouraging me to introduce Rawls's primary goods measure in greater detail.
}
However, unlike the simple measure of income and wealth, Rawls's conception of resources is broad and inclusive. The primary goods consist of both \emph{material} resources (such as income and wealth) and also \emph{immaterial} resources (such as opportunities, rights, and liberties, among others)~\citep{rawls_theory_1971}. Rawls conceives the primary goods measure in this way (in part) to achieve greater sensitivity to the heterogeneous social and political conditions that affect people's well-being in society. This is because (as above) these heterogeneities may not be readily expressed in the quantifiable terms of income and wealth alone.

While Rawls's primary goods measure includes much more than income and wealth, the measure may nevertheless be insensitive in other ways. It may not, for example, adequately attend to disparities in well-being that are not constituted by a lack of resources (either material or immaterial) but rather by social and political relations between people, as we will see later on. Although it is conceived broadly and inclusively, the primary goods measure is still ultimately ``resourcist'' in its conception (i.e., constituted by resources), and so the capability theorists argue that it is ultimately inadequate for the purpose of remediating social injustice. Capability theorists object not so much with which (or how many) resources are included in the primary goods measure, but with resources \emph{themselves}, as a type of benefit to be looked at as a measure of justice. On their view, we can (and should) measure another type of benefit: one that achieves greater sensitivity to people's heterogeneous social and political conditions, thereby better encompassing what contributes to people's well-being in society. Capability theorists like Sen defend a measure of people's \emph{capabilities}.

To foreshadow, I will argue that measures from ML \emph{resemble} resource measures such as Rawls's, and that this resemblance makes the resourcist approach amenable to operationalization in ML systems. 
At the same time, I will argue that this resemblance opens measures from ML to similar critiques from the capability perspective. Specifically, it opens these measures to the critique that they are \emph{insensitive} to the heterogeneous social and political conditions that affect people's well-being in society.
Note that the relation between resource measures and those from ML is one of resemblance, and not of equivalence. This is because Rawls's primary goods measure is much broader and more inclusive than the measures appearing in resource allocation problems from \ac{fair ML}. An equivalence relation would be more appropriately drawn between measures from ML and the simple resource measure of income and wealth. That is, measures from ML tend to consist of (at most) a very small subset of Rawls's primary goods, which are conceived broadly and inclusively. So, if the capability theorists' critiques of the primary goods measure land with any force, we might expect similar critiques to land much more forcefully, when leveled against measures much more limited than Rawls's.

\subsection{Two Criteria for a Measure of Justice}
\label{section:2.2}

I will establish a theoretical resemblance between measures from ML and resource measures from theories of distributive justice. To do this, I first introduce two important criteria for a measure of justice, which are at the center of the resources-versus-capabilities debate: the measure's sensitivity to personal heterogeneities and societal context (which I refer to as the \textbf{sensitivity criterion}) and the measure's public legibility and verifiability (or the \textbf{publicity criterion}). Introducing these criteria early on will help us to outline the contours of the debate, and to later draw the resemblance between resource measures and their operationalized analogues in ML systems. As I introduce these criteria, the reader may find it helpful to keep in mind the overarching argumentative point: that measures from ML \emph{resemble} resource measures, and are therefore open to similar critiques from the capability perspective.

\subsubsection{The Sensitivity Criterion}
\label{section:2.2.1}

Should a measure of justice be sensitive to people's heterogeneities and societal context? That is, should a measure take into consideration differences between people, in terms of their physiological and psychological functioning as well as their social and political conditions? Sen and other capability theorists argue that because a person's well-being depends not only on the resources they have in their possession, but also on their heterogeneous conditions, a measure of justice must be sensitive to these conditions.

To appreciate what it means for a measure to be sensitive, consider one way that resource measures are argued to be \emph{insensitive}. According to capability theorists, resource measures are not adequately sensitive to what contributes to the well-being of disabled people~\citep{brighouse_justifying_2010,rawls_equality_1979,brighouse_what_2010}. This is because these measures do not attend to the complex interplay between bodily functioning and societal context, from which disability arises.\footnote{
 In disability scholarship, the \emph{social model} of disability makes a conceptual and terminological distinction between ``impairment'' in bodily functioning and ``disability''. Impairment is characterized by some limitation of physiological or psychological functioning, whereas disability is characterized by social or political exclusion perpetrated on the basis of that impairment. Impairment alone is not disabling. Rather, disability arises when a society---built to accommodate ``normal'' people who lack ``abnormal'' impairments---ends up excluding (intentionally or not) people with impairments from full social and political participation~\citep{wasserman_disability:_2016, shakespeare_social_2013}.
 While the social model has been enormously influential, the disability-impairment distinction has been critiqued in the Disability Rights and Disability Pride movements, as well as in more recent scholarship. According to these critiques, the distinction between ``disability'' and ``impairment'' is untenable, and possibly circular. If disability is understood as societal exclusion perpetrated on the basis of ``abnormal'' impairment, then there must some characteristics that distinguish people who are ``impaired'' (i.e., limited in their in bodily functioning) from people who are not. But, it's unclear what those characteristics are, exactly, since impairment may itself be socially determined or specific to a person's historical and cultural context~\citep{tremain_government_2001}. An alternative view, known as the \emph{minority model} of disability, proposes that what it means for a person to be ``disabled'' is to be viewed as such by their society, and to be socially subordinated as a minority within that society. That is, to be disabled is to be disadvantaged (along some dimension) in comparison to people who are non-disabled, without positing impairment in bodily functioning as the basis for that disadvantage~\citep{barnes_minority_2016}. I thank Sally Haslanger for suggesting these references.
} For example, two people may be equal in their allotment of resources, and thus apparently equal in their well-being, according to the simple measure of income and wealth. But, if one person has a disability that is monetarily expensive (due to structural barriers, a lack of accommodation, or outright social exclusion), then the two may in fact be unequal in their well-being~\citep{rawls_equality_1979}. This critique is easy enough to grasp when considering the simple measure of income and wealth, but the capability theorists also level it against Rawls's primary goods measure, broadly and inclusively conceived. Equalizing resources of any sort, they argue, will inadequately remediate social injustices that arise from people's heterogeneous social and political conditions. Equality will be achieved in name only. 

But here it may be asked: why not adjust the rule instead of the measure? That is, couldn't an \emph{unequal} reallocation of resources---whereby disabled people receive \emph{more} resources---adequately remediate these social injustices?\footnote{
    I thank Harini Suresh for encouraging me to clarify this point.
}
This is a plausible idea. But, it only moves the problem out of view temporarily, and the capability theorists ultimately reject it.\footnote{
    As with equalizing income, unequal redistribution of resources toward disabled people---although well-intentioned---may yet be an inadequate remediator of social injustice, in two ways. First, it may not allow people with disabilities to bypass long-standing structural barriers when doing so requires accommodations that exceed their allotted resource bonus, or when doing so is blocked by social prejudice (e.g., as when workplace modifications are deemed ``unreasonable'' by an unaccommodating employer). Second, reallocation of resources toward disabled people may disrespect their personhood, or be indignifying, by suggesting that they are somehow pitiful, or that their disabilities are quantifiable losses rectifiable by resource compensation, thereby entrenching a subordinate social status relative to their non-disabled peers~\citep{anderson_what_1999}.
}
As above, reallocating resources can inadequately encompass well-being---for people with disabilities, especially---but also for \emph{anyone} who experiences discrimination, structural barriers, psychosocial harms, and other forms of non-monetary, less-readily-quantifiable disadvantage.
Structural and psychosocial injustices---including  \emph{de facto} segregation, discourse inequality, stigma, and shunning---are prevasive, but they are not likely remediable by reallocations of resources alone~\citep{brighouse_justifying_2010}.
This is because these injustices are not constituted by a inequitable distributions of resources \emph{across} people, but rather by social and political relations \emph{between} people.

Here, I note (briefly) that \ac{fair ML} often posits a distinction similar to the one just described. That is, a distinction is posited between algorithmic unfairness which \emph{can be} mitigated via resource reallocation versus unfairness which \emph{cannot} because it arises from social and political relations between people, and not from an inequitable distribution of resources alone. In the \ac{fair ML} literature, this is known as the distinction between \emph{allocative harms} and \emph{representational harms}~\citep{crawford_trouble_2017, barocas_fairness_2019}.
I will postpone a fuller introduction to this distinction until later on, when I discuss how \ac{fair ML} researchers have sometimes conflated allocation and representation when operationalizing theories of distributive justice~\sectionlabel{section:4.2}.
It suffices to note here that measures which are sensitive to heterogeneous political and social conditions (namely, capability measures) are conceived to better address harms of the second sort (namely, harms of representation, not of allocation).
But, if sensitive measures do a better job of addressing these harms, it may be wondered: why not select a sensitive measure? Because, the sensitivity criterion is in tension with another: publicity. 

\subsubsection{The Publicity Criterion}
\label{section:2.2.2}

At the level of a society's systems and institutions, the information required to make a claim of injustice must be publicly legible and verifiable~\citep{brighouse_measuring_2010}.
That is, the measure must be transparent (not murky) and accessible to all.
This is because, if this information were not publicly legible and verifiable, the legitimacy and stability of the systems and institutions regulated by such a measure may be suspect or indeterminate~\citep{brighouse_justifying_2010}. 
Inconsistent and/or inscrutable measures may themselves become sources of social injustice when they stymie public oversight and accountability~\citep{brighouse_two_2010}.
Here, resourcists and capability theorists agree: the right measure of justice must be publicly legible and verifiable. They disagree, however, on whether both the publicity and sensitivity criteria can be satisfied at the same time.

Resourcists argue that, for a measure to satisfy the publicity criterion, it cannot also be sensitive to people's heterogeneities and societal context. Only if a measure is insensitive will it provide a \emph{public standard} of measurement: one that is consistently applicable and legible across diverse groups of people and heterogeneous social and political conditions~\citep{brighouse_equal_2010}.\footnote{
 For example, in the physical sciences, the International System of Units, or metric system, is one such public standard of measurement. It allows scientists to understand, share, and compare their findings in a consistent fashion across different contexts, helping to ensure that those findings are publicly legible and, when necessary, verifiable.
} Sensitive measures will fail to provide a public standard, resourcists argue, for at least two reasons. First, sensitive measures likely require large quantities of information that may be difficult or undesirable to acquire in practice (perhaps due to parsimony or concerns about individuals' privacy). This makes them difficult to verify in a public fashion~\citep{robeyns_capability_2016}.
Second, even if these information demands could be met, sensitive measures (by definition) vary according to people's heterogeneities and societal context. This potentially renders indeterminate (or murky) comparisons between different people in different contexts~\citep{brighouse_critique_2010}. 

For comparison, consider again the simple resource measure of income and wealth. Although this monetary measure may be insensitive to people's heterogeneities and societal contexts, it is at least \emph{consistently} measurable across people and contexts. That is, while income and wealth may not encompass the \emph{entirety} of a person's well-being, it can at least provide a consistently measurable indication of well-being across heterogeneous social and political conditions. Although capability theorists and resourcists agree that a measure must be publicly legible and verifiable, they disagree on whether a measure can satisfy both criteria at the same time. Capability theorists argue that a measure can satisfy both, as we will see later on, whereas resourcists argue otherwise.

\subsection{Sensitivity vs. Publicity}
\label{section:2.3}

In the resources-versus-capabilities debate, there exists a tension between a measure's sensitivity and its publicity. On the one hand, measures that are insensitive to people's heterogeneous conditions may overlook, encode biases against, or perpetuate social injustice against people whose well-being matters in evaluations of justice and fairness. On the other hand, measures that are sensitive to people's heterogeneous conditions may obscure injustices at the level of systems and institutions, if those measures are publicly illegible, inconsistently applicable, or require large quantities of information that are prohibitively costly or undesirable to acquire (perhaps due to overriding concerns about individuals' privacy).

So far, I have said a lot about the philosophical criteria for a measure of justice, and only a little about ML. However, I claim that the tension between publicity and sensitivity---between resources and capabilities---from the prior discussion carries over to the design and development of ML systems. The prior discussion begins to suggest how it is that resource measures and measures from ML are alike: they are both publicly legible and verifiable, but may be insensitive to people's heterogeneous social and political conditions.
In the next section, I draw this resemblance much more explicitly. While the resemblance may seem readily apparent, simply from observing which distributive theories \ac{fair ML} has picked out for operationalization, I will argue that it goes deeper than theory selection, or a superficial similarity between ``resource'' terminology. At bottom, resource measures and measures from ML share similar \emph{characteristics} that contribute to their satisfaction (or not) of the publicity and sensitivity criteria.
\section{The Resourcist Approach to Measuring Justice} 
\label{section:3}

In this section, I first define what resources are, as they are understood in both the distributive justice and \ac{fair ML}, and provide some examples of each. Then, to establish the resemblance between resource measures and measures from ML, I present two key characteristics shared by the measures appearing in both literatures. Based on these shared characteristics, I argue that measures from ML satisfy the publicity criterion, but fail to satisfy the sensitivity criterion. In upcoming sections, I show how this failure can occur in practice, by discussing two real-world case studies in which the measures selected for mitigating algorithmic unfairness end up entrenching and reproducing social injustice---just as the capability theorists argue resource measures do.

\subsection{What Are Resources?}

Recall that the right measure of justice is intended to encompass what it is that contributes to people's well-being within their society. The resourcist approach aims to do this by looking primarily at the resources people have in their possession. A person's \textbf{resources} are defined as the things that they can possess or use, in the broadest sense~\citep{long_egalitarianism_2019}.

In theories of distributive justice, resources can be \emph{material} (such as bundles of various goods, income, and wealth) as well as \emph{immaterial} (such as services, opportunities, liberties, basic political rights)~\citep{cohen_currency_2011}. As I previously noted, Rawls's primary goods measure conceives of resources in this broad and inclusive way, to better encompass what it is that contributes to people's well-being. Acquiring more resources usually improves people's well-being. For example, people can usually use their income to buy a house to attain shelter, food to attain sustenance, health insurance to attain care, a train ticket to attain mobility, a computer to attain internet access, and so on---all of which tend to yield improvements to a person's quality of life. Because people can usually use resources in this way, resourcists argue that well-being is more-or-less encompassed by the resources people have in their possession. Whoever has more resources at their disposal is generally better-off, disregarding other personal heterogeneities.

In \ac{fair ML}, the resources appearing in ``resource allocation'' problems have lacked a consistent and agreed-upon definition. In many cases from the literature, the resources under consideration are material: loans (to be fairly allocated across loan applicants)~\citep{jabbari_fairness_2017, hu_welfare_2018}, power generators (to be fairly allocated across homes)~\citep{donahue_fairness_2020}, and so on. In other cases, the resources are immaterial: university admissions (to be fairly allocated across students)~\citep{dwork_fairness_2011}, salary opportunities (to be equalized across employees)~\citep{heidari_fairness_2018}, and so on.
In the vast majority of cases, however, what is looked at as a measure of fairness is some computational artifact of the ML model itself: the model's predictive accuracy, or false positive and negative rates, for different groups of people~\citep{narayanan_21_2018}.
These cases are peculiar. What is the ontological status of these computational artifacts?
In one sense, they are simply numerical quantities: abstract representations of statistical concepts like accuracy, error, utility, likelihood, and risk.
In another sense, the numerical values obtained by these artifacts are constrained by algorithmic optimization procedures, which are both constrained mathematically and by real-world (i.e., finite) hardware resources, such as those of a computer's processing units.
Does this suggest that these computational artifacts are \emph{themselves} ``resourcist'' in their conception? That is, are they things that people can possess or use, in the broadest sense?
This is not entirely clear. But, what is clear is that, although they are ``merely'' numerical quantities, these computational artifacts are often referred to in allocative terms.
That is, artifacts like predictive accuracy are referred to as being ``allocated'' across people qua the group of which they are a member.\footnote{
    I thank Reuben Binns for suggesting this language.
}
For example, researchers describe ``fair allocation of predictions'' across groups of people~\citep{zliobaite_survey_2015} and refer to ``allocational harms'' that are understood as disparities in a model's accuracy for different groups of people~\citep{de-arteaga_bias_2019, davidson_racial_2019}.
In another example, a model's predictive accuracy is \emph{explicitly} characterized as a ``resource'' to be fairly allocated via a distributive rule~\citep{hashimoto_fairness_2018}. 
These examples suggest that the ``resources'' often appearing in \ac{fair ML} are conceptualized---not only as material or immaterial---but also as uniquely \emph{computational} in some sense.
Here, I do not attempt to give a complete ontology of these ``computational resources''. 
It is enough to note two things. First, computational artifacts like predictive accuracy share important characteristics with resource measures from distributive justice, as I will discuss in the next section. Second, when these artifacts are ``allocated'' across different groups of people, they have the power to improve or diminish people's well-being within their society~\citep{angwin_machine_2016, buolamwini_gender_2018, dastin_amazon_2018}.
It is in this latter sense that computational artifacts may be ``posessed'' or ``used'' in different ways, by different people, depending on their heterogeneous social and political conditions.

In all of the above cases, both from distributive justice and from \ac{fair ML}, the resources allocated are generally thought to improve the well-being of people who receive or possess them (be it through direct acquisition in the real world, or through interaction with a ML system). That is, whoever has more resources (be they material, immaterial, or computational) is generally assumed to be better-off.

\subsection{Two Characteristics of Resource Measures}

Previously, I introduced two criteria for a measure of justice (sensitivity and publicity) and discussed at a high level how it is that resource measures do or do not satisfy those criteria. However, I did not discuss at a low level why that is exactly. That is, I left out the specific \emph{characteristics} of resource measures that contribute to their satisfaction of the publicity criterion, but not the sensitivity criterion. In this section, I introduce two key characteristics of a measure of justice, and argue that resource measures and measures from ML align on both.\footnote{
 I do not claim that these characteristics exhaust all possible similarities and differences between the measures under consideration. Indeed, there is a third key characteristic: whether the measure is \emph{subjective} or \emph{objective}. That is, whether what is being measured exists ``in the world,'' outside of people's subjective mental states, such as their mental states of happiness or preference satisfaction~\citep{brighouse_justifying_2010}. Note that ``objective'' here refers to existing ``mind-independently'' and not to scientific objectivity, or observing impartially and without bias. However, because the objective-subjective distinction is introduced to distinguish objective measures (such as resources and capabilities) from subjective measures (such as welfare and utility), I do not discuss it at length, since all of the measures presently under consideration (resources, capabilities, and measures from ML) are ``objective'' in this sense.
} This alignment forms the basis of my resemblance claim between the two measures.

The two key characteristics are: (1) the measure can be expressed as a single-valued quantity, rather than multi-valued quantities or qualities, and (2) what is being measured (namely, a resource) is a ``means to an end,'' rather than an ``end in itself.'' To better understand these characteristics, I discuss them as they apply to both resource measures and measures from ML.
For easy reference, \autoref{tab:measuretable} shows how these measures align on the key two characteristics, and suggests how it is that these characteristics do or do not contribute to satisfaction of the sensitivity and publicity criteria. Additionally, this table foreshadows how capability measures differ in their characteristics and satisfaction of the criteria.

\subsubsection{Resource Measures Can Be Single-Valued}

Resources are things that people can possess or use, in the broadest sense. People can possess or use many things---income and wealth, material goods, opportunities, rights and liberties---any one or all of which may be looked at as a measure of justice. But, not all resources are the same. Some have characteristics that make them more or less amenable to certain types of measurement (quantitative or otherwise) and to satisfying the publicity and sensitivity criteria for a measure of justice.

As I previously noted, Rawls specifies his primary goods measure broadly and inclusively (in part) because different resources can be called upon to play different roles when detecting and remediating social injustice. Of particular importance, some resource measures can be---but do not \emph{need} to be---expressed as single-valued quantities. This, it turns out, is a very useful characteristic. For example, monetary resources like income and wealth are particularly useful because they can be expressed in singular, quantitative terms (i.e., as the numerical quantity of income or wealth a person has in their possession). When included as part of a measure of justice, single-valued quantities are useful because they permit unambiguous comparisons, orderings, and interval rankings of the people whose well-being they aim to encompass. To know, for example, who is better-off when using a monetary measure like income, we simply need to look up and compare the numerical quantities of income that people have in their possession. Thus, according to such a single-valued quantitative measure, it is always clear to us where people stand in relation to one another, even across heterogeneous social and political conditions. This is a win for the measure's publicity, but simultaneously, it is a loss for the measure's sensitivity, for reasons that we have already seen.

By contrast, measures that are multi-valued (i.e., incorporating more than a single variable) or heterogeneous (i.e., mixing qualitative and quantitative data) are much less likely to permit unambiguous comparisons, orderings, and interval rankings of the sort achievable via single-valued quantities.
This is because the information admissible to these measures includes \emph{more than} a single-valued quantity. It may, for example, include qualitative textual data, or it may include multiple quantities such that no single quantity can be called upon to establish a strict interval ranking.
These sorts of measures are more common to mixed-methods research, ethnography, narrative inquiry, grounded theory, or other social science methodologies.

Here, it is important to state explicitly: resource measures \emph{can be} single-valued, but they are not \emph{necessarily} single-valued.\footnote{
    I thank Gabriel Karger for encouraging me to clarify this point.
}
We could, for example, formulate a quantitative multi-valued resource measure, so that it does not strongly prioritize any constituent single-valued quantity, such as income and wealth. Indeed, we will see later on how the capability approach (in some of its variants) proposes to do exactly that. But, while we can formulate such a multi-valued resource measure, we will at the same time be leaving behind one of the primary reasons resources are useful in evaluations of justice and fairness: when expressed as single-valued quantities, resource measures permit unambiguous comparisons and rankings. Being able to make these comparisons and rankings is crucial to the success of Rawls's distributive principles.\footnote{
 It is worth clarifying in detail how, exactly, capability theorists argue that Rawls is committed to single-valued quantitative measures. Rawls's measure of primary goods is conceptualized broadly to include both material resources (income and wealth) and immaterial resources (opportunities, rights, liberties, and others). This suggests that the primary goods measure is \emph{not} single-valued, but rather multi-valued. Certainly, the primary goods include resources other than income and wealth alone. And yet, the capability theorists still critique primary goods for providing a singular quantitative standard. For example, Martha~\cite{nussbaum_capabilities_2009} says that Rawls is committed to ``measuring relative social positions in a single and linear way, with reference to income and wealth alone.'' Are capability theorists like Nussbaum flatly misunderstanding Rawls's view, interpreting it uncharitably, or is there something more going on here? The last question (and some might argue the middle one too) can be answered affirmatively. The reason capability theorists critique the primary goods measure as ``single-valued'' has to do with the distributive rule that Rawls pairs it with, what is known as the Difference Principle. As it turns out, \ac{fair ML} researchers have picked out Rawls's Difference Principle for operationalization in ML systems, so I will postpone a full introduction until later on~\sectionlabel{section:4.2}. It is enough to note here that income and wealth (as constituents of the primary goods measure) are called upon to play a central role in the successful operation of Rawls's Difference Principle. According to~\cite{nussbaum_capabilities_2009}, Rawls is committed to measuring relative social positions ``with reference to income and wealth alone'' because he ``attaches considerable importance to the ability to rank in a definite and unilinear way who is well-off and not well-off... if the measures were [multi-valued] and heterogeneous, then it would be unclear who is least well-off, and the whole argument for the Difference Principle would be thrown into jeopardy.'' Some defenders of Rawls's view, such as Thomas~\cite{brighouse_critique_2010}, agree with Nussbaum on this point: the Difference Principle requires ``not merely a partial ordinal ranking, but a complete interval ranking.'' And such a ranking would not be possible if the measure were multi-valued and heterogeneous.
}
Likewise, it is just as crucial (if not more so) to the successful operation of algorithms appearing in \ac{fair ML} and machine learning systems. If the distributive and allocational rules that regulate these systems were instead combined with multi-valued and/or heterogeneous measures---resulting in potentially ambiguous or murky comparisons and rankings---then their entire operation may be thrown into jeopardy (see footnote 13).

\bgroup
\def\arraystretch{1.5}
\begin{table}[]
\centering
\begin{tabular}{c|p{28mm}|p{28mm}?p{28mm}|p{28mm}|}
\hhline{~----}
& \multicolumn{2}{c?}{\textbf{Characteristics}} & \multicolumn{2}{c|}{\textbf{Criteria}} \\ 
\hhline{~----}
& can be a single-valued quantity? & measure a ``means to an end''? & are sensitive to heterogeneities? & are publicly legible \& verifiable? \\ \hhline{-----}
\multicolumn{1}{|c|}{Resource measures}   & {\cellcolor{ForestGreen!15}}Yes & {\cellcolor{ForestGreen!15}}Yes & {\cellcolor{Mahogany!15}}No & {\cellcolor{ForestGreen!15}}Yes \\ \hhline{-----}
\multicolumn{1}{|c|}{Measures from ML}         & {\cellcolor{ForestGreen!15}}Yes & {\cellcolor{ForestGreen!15}}Yes & {\cellcolor{Mahogany!15}}No & {\cellcolor{ForestGreen!15}}Yes \\ \hhline{-----}
\multicolumn{1}{|c|}{Capability measures} & {\cellcolor{Mahogany!15}}No & {\cellcolor{Mahogany!15}}No & {\cellcolor{ForestGreen!15}}Yes & {\cellcolor{gray!15}}Yes? \\ \hhline{-----}
\end{tabular}
\caption{\label{tab:measuretable}The measures, their characteristics, and whether or not they satisfy the publicity and sensitivity criteria. Resource measures and measures from ML are aligned on their characteristics and their satisfaction of the criteria. (Although not discussed at length in this paper, responses to these questions regarding the welfare measures would read left-to-right as Yes, No, Yes, No.)
}
\end{table}
\egroup

\subsubsection{Resources Are Means, Not Ends}

In the resources-versus-capabilities debate, an important distinction is posited between things that are a ``means to an end'' and things that are an ``end in themselves.'' A means to an end is something that people can use to achieve some other thing: another means or, eventually, something that is an end in itself. In short, ends are things that are \emph{ultimately} valuable to people, and means are merely a way of getting to them.

\noindent
The means-end distinction is posited to argue that resources are a means to well-being, which is an end in itself~\citep{long_egalitarianism_2019}. Recall that two people may possess equal resources (such as income and wealth, or primary goods) and yet still be unequal in their well-being, if one lives under heterogeneous social and political conditions that diminish their well-being. Thus, it is argued, what contributes to people's well-being must consist of something \emph{other than} (or in addition to) resources alone~\citep{rawls_equality_1979}. Resources may be a means to achieving well-being, but resources are not ultimately valuable to people, as an end in themselves. What people \emph{ultimately} value is their well-being, not the resources they use to achieve it.

Similarly, the things that are ultimately valuable to people with respect to ML systems may not be what \ac{fair ML} researchers have so far attempted to measure~\citep{thomas_problem_2019}. For example, \ac{fair ML} researchers have so far been concerned with the accuracy of a ML model, or similar computational artifacts. A model that is less accurate for some people, and more for others, may seem flatly unfair to the former. Accordingly, researchers may adjust their model's optimization procedure so that it achieves accuracies more evenly balanced across these different groups of people. When accuracy is roughly equalized across people, \ac{fair ML} researchers might think that their job is done; fairness accomplished. But, as with the resources described above, accuracy might not be what people \emph{ultimately} value when interacting with ML systems.
This is not to suggest that people don't value accuracy \emph{at all}, but that what people ultimately value includes \emph{more than} accuracy alone.
Accuracy may be a means or a ``proxy'' for what people value with respect to ML systems, but it may not be an end in itself. Thus, an inordinate focus on accuracy as the sole measure of the fairness or justness of these systems may elide other important ways in which people's well-being is improved or diminished by them, as I will discuss in the upcoming case studies.

\subsection{Resource Measures: Publicity, Not Sensitivity}

Measures commonly used in \ac{fair ML} resemble resource measures from distributive justice to the extent that they align on at least two key characteristics: (1) these measures can be expressed as single-valued quantities that permit unambiguous comparisons and rankings, and (2) what is being measured (namely, a resource of some sort) is a means to an end, rather than an end in itself. Taken together, these two characteristics contribute to measures that satisfy the publicity criterion, but fail to satisfy the sensitivity criterion.

Consider publicity. Because resource measures and measures from ML can be expressed as single-valued quantities, they provide a \emph{public standard} of measurement that is consistent and practicable across heterogeneous social and political conditions. This is because it is easy to measure people's relative social positions (by simply looking up their income, or some other single-valued quantity). Single-valued quantities also permit unambiguous comparisons and interval rankings (by simply comparing these singular values). In this way, these measures are publicly legible and verifiable. In ML, for example, the accuracy of different models for different groups of people can be compared consistently on benchmark datasets, allowing researchers to share their findings in a publicly legible (and if necessary, verifiable) fashion.

Consider sensitivity. Because resource measures and those from ML focus on means (i.e., material, immaterial, or computational resources) rather than ends in themselves (i.e., things that people ultimately value), without considering how the two can come apart, these measures can be insensitive to heterogeneous social and political conditions that are relevant to evaluations of algorithmic fairness or distributive justice. This is because not all algorithmic unfairness (or social injustice) is readily expressed in terms of quantifiable resource deficits: some unfairness or injustice is structural and psychosocial, arising from social and political relations between people.
Thus, the capability theorists argue that resource measures achieve their publicity at the expense of sensitivity.
According to Elizabeth~\cite{brighouse_justifying_2010}, resource measures pass over ``structural and psychosocial injustices that interfere with individuals' functioning as equals, although [these injustices] are neither constituted nor remediated by distributions of resources.''
In the next section, I discuss two real-world cases wherein structural and psychosocial injustices go undetected and unremediated by measures from ML the exhibit these characteristics.

\section{Case Studies from Machine Learning}
\label{section:4}

To further substantiate the resemblance between resource measures and measures from ML, and to illustrate the differences between the resourcist and capability approaches when operationalized in ML systems, I discuss two case studies. Each case concerns a distributive principle that specifies a seemingly fair distributive rule, but---by also specifying a resource measure---the principle fails to adequately remediate structural and psychosocial injustice. 

In the first case, I discuss how computational resource measures used in algorithmic hiring can encode and reproduce historic and ongoing employment discrimination, specifically against disabled people. In the second case, I discuss how the measures used in operationalized analogues of Rawls's theory can subject racial minority groups to stereotype threat, discourse inequality, or other psychosocial harms. In both cases, I argue that the specified measures exhibit the two key characteristics from the the previous section, such that they satisfy the publicity criterion but fail to satisfy the sensitivity criterion. I contextualize and motivate each case with an introductory discussion of its political and historical significance.

\subsection{Algorithmic Antidiscrimination}
\label{section:4.1}

Discrimination has long pervaded employment. From the Civil Rights Act of 1964, to the Americans with Disabilities Act of 1990 (ADA), the long history of antidiscrimination law makes apparent that employment opportunities have rarely, if ever, been fairly distributed in the United States~\citep{united_states_americans_1990}. Even long after the passage of these laws---which grant employment protections to disability, gender, and racial minority groups, among others---employers have continued to discriminate, often motivated by cognitive bias, if not explicit prejudice~\citep{ameri_disability_2015, bertrand_are_2003, cole_recruiters_2007}.

Against this backdrop of historic and ongoing discrimination, proponents of algorithmic decision-making argue that introducing ML into hiring decisions could reduce bias against these legally protected groups by enforcing fairness constraints on the pool of job candidates under consideration.\footnote{
 Companies like LinkedIn, Indeed, and Pymetrics use hiring algorithms to place and optimize job advertisements appealing to target demographics, notify jobseekers about positions of potential interest, entice out-of-work candidates to rejoin the job market, administer pre-employment application tests, and recommend promising candidates to recruiters~\citep{bogen_help_2018,carey_how_2016,silverman_algorithm_2015}.
} Algorithmic decision-making, these proponents argue, affords an opportunity to remediate employment discrimination by removing human prejudice and cognitive bias from the decision-making process. At the same time, critics argue that introducing algorithms into hiring decisions could silently encode, entrench, and reproduce human biases against legally protected groups, by recommending only those candidates whose characteristics match high-performing employees belonging to the already dominant groups (typically comprised of young, white, non-disabled, cis-gender men).\footnote{
 For example, in 2018, Amazon abandoned its algorithmic review and recommendation of job applicants when their ML model (trained on 10 years of resumes submitted to the company) was found to be unfairly biased against women~\citep{dastin_amazon_2018}. As early as 2016, the Equal Employment Opportunity Commission (the United States agency responsible for enforcing employment antidiscrimination laws) raised concerns about the potential for hiring algorithms to reproduce biases against women and disabled people, if they happened to have irregular work habits or a pattern of absences, due to child-care obligations or healthcare needs~\citep{elejalde-ruiz_alexia_end_2018, equal_employment_opportunity_commission_use_2016}.
} 

While all legally protected groups have faced longstanding and on-going discrimination in the workplace, studies show that disabled people have been especially disadvantaged.\footnote{
 In 2018, nearly 30 years after the passage of the ADA in the United States, the employment-population ratio (i.e., the proportion of the population that is employed) was 19\% among adults with a disability. By contrast, the employment-population ratio for adults without a disability was 66\%~\citep{labor_statistics_persons_2018}. The disparity between disabled and non-disabled populations represented one of the largest, indicating substantial barriers to employment for disabled people. For comparison, in 2018, the employment-population ratio was 56\% among adult women versus 69\% among adult men, 63\% among black men versus 67\% among white men, and 58\% among black women versus 56\% among white women.
} In addition to the barrier of identity-based prejudice, people with disabilities face a unique barrier to employment: the barrier of a work environment not built from the outset to accommodate the many ways of moving and working in society. Today, after decades of disability rights activism and advocacy for equal access to work environments, disabled people face another, analogous barrier: an information technology environment not built from the outset to accommodate the many ways of being and working online. As hiring decisions become increasingly mediated via online ML systems, the long history of disability employment discrimination casts doubt on claims that this new information technology environment will remediate---rather than entrench and reproduce---employment discrimination against people with disabilities.

\subsubsection{The Case of Disability Hiring}
\label{section:4.1.1}

Consider jobseekers who browse for employment opportunities on online job platforms, such as LinkedIn, Indeed, or Pymetrics. Hiring algorithms deployed on these job platforms search through resumes for promising candidates (i.e., those fitting qualifications provided by company recruiters) and send them \emph{pre-employment application tests} (i.e., timed tasks intended to test a candidate's relevant skills), administered through the job platform's website~\citep{bogen_help_2018, carey_how_2016}. Suppose, charitably, that these algorithms are designed to take into consideration any available demographic information (such as whether a candidate belongs to a dominant or legally protected group) and follow a rule for allocating application tests based on a bias-mitigating conception of algorithmic fairness. In other words, suppose that the hiring algorithm is optimized to distribute a ``fair'' number of application tests (employment opportunities) across demographic groups, which it does by allocating fewer tests to dominant group candidates and more to legally protected group candidates, subject to the demands of fairness. 

Because the distributive rule accounts for disparities between dominant and legally protected candidate groups (by targeting the latter with more opportunities to complete application tests), such a reallocation of pre-employment application tests may seem sufficient for achieving algorithmic fairness, in this context. But what about the fairness of the measure itself: the application tests?

Not implausibly, these tests could encode biases against legally protected groups of people. For example, blind jobseekers who use screen readers may be relatively disadvantaged if the tests themselves are not built for screen reader compatibility, as they typically are not.\footnote{
     Screen readers are a type of assistive technology that convey visual information through non-visual means (such as text-to-speech or braille) in order to make internet browsing accessible to blind people.
      In 2019, Domino's Pizza, Inc. brought a case to the United States Supreme Court, arguing that their company website should not be required to comply with Title III of the ADA (i.e., arguing that it does not need to be made accessible to blind customers who use screen readers)~\citep{barnes_dominos_2019}. Title III of the ADA requires that ``[n]o individual shall be discriminated against on the basis of disability in the full and equal enjoyment of the goods, services, facilities, privileges, advantages, or accommodations of any place of public accommodation''~\citep{united_states_americans_1990}. The Supreme Court declined to hear the case.
} That is, two equally-qualified candidates may receive the same application test (and thus have an apparently equal opportunity for employment), but one may be unfairly disadvantaged, if the test is in some way inaccessible to jobseekers who use screen readers.\footnote{
 For example, if the application test is not designed in accordance with the World Wide Web Consortium's Web Content Accessibility Guidelines (WCAG), people who use screen readers may not be able to access certain content. If the application test uses a candidate's time-to-completion in assessments of their performance, screen reader users may be at a competitive disadvantage, since website navigation may take longer with the use of a screen reader than without~\citep{trewin_ai_2018}. However, when reading text, screen reader users can be significantly faster than people who read visually.
} Allocating more pre-employment application tests toward screen reader users will thus inadequately remediate employment discrimination, if the design of those tests is based on a biased conception of a job candidate's functioning. Merely measuring the quantity of application tests a candidate receives---without providing opportunities for recourse via a human hiring agent, for requesting accessibility accommodations, or for contesting automated decisions---will therefore be insensitive to people's heterogeneous ways of being and working online.
Furthermore, seeking recourse, requesting accommodation, or initiating contestation within an algorithmic hiring system is likely to be difficult when that system's behavior is largely opaque, both to jobseekers (who may not notice when and how they are unfairly disadvantaged), and to employers (who may have little incentive to scrutinize the system's recommendations).
While human hiring agents are also fallible, their behavior is at least scrutable, and their decisions can therefore be contested and influenced through dialogue and interaction.

Observe here that measuring the quantity of pre-employment application tests a candidate receives exhibits two key characteristics of a resource measure: (1) it is a single-valued quantity that is easy to measure, compare, and rank (by simply looking up the number of tests allocated to jobseekers belonging to different demographic groups), and (2) it is a means or ``proxy'' for what actually contributes to a jobseeker's well-being (namely, real opportunities to achieve employment, free from structural or technological barriers). Thus, such a measure satisfies the publicity criterion, but fails to satisfy the sensitivity criterion. Although consistently measurable across different jobseekers, merely receiving the opportunity to complete a pre-employment application test is \emph{insensitive} to whether or not a candidate is capable of completing it, given their particular way of being and working online.\footnote{
 Furthermore, quantitatively measuring the allocation of any single computational resource or technological artifact (such as the quantity of tests allocated or otherwise) can in principle encode biases in the same way. Because these artifacts are designed with a particular conception of human functioning in mind, they may not be sensitive to the context-specific ways in which people are capable or incapable of using them. Thus, receiving more (or fewer) artifacts does not necessarily improve (or diminish) a person's well-being. Adherents to the resourcist approach might counter-argue that computational resources or technological artifacts of this sort are not the right resources to look at, as a measure of fairness or justice for ML systems. That is, there could be other resources, say, those existing \emph{outside} of ML systems, which are more appropriate to evaluations of justice or fairness, in this context. This may be correct. However, if correct, the standard capability critique of non-computational resources would then apply.
}

\subsection{Operationalizing Rawls's Difference Principle}
\label{section:4.2}

Throughout, I have noted that \ac{fair ML} researchers have sought to operationalize theories of distributive justice in ML systems, in particular those belonging to John Rawls's resourcist approach. Here, I discuss one such case in detail.
This case is significant to the inquiry for three reasons: (1) it exhibits a straightforward attempt by \ac{fair ML} researchers to operationalize one of Rawls's distributive principles, namely the Difference Principle, (2) in the operationalization, the ML model's predictive accuracy is selected as the measure and is \emph{explicitly} characterized as the ``resource'' to be allocated across people via the rule, and (3) the operationalization is carried out to address the problem of \emph{representation disparity} in ML models.
The reader can likely grasp why the first two reasons are significant, but perhaps not the third, which requires further explanation.

The problem of representation disparity occurs when an ML model achieves high overall accuracy for a population, but low accuracy for a subgroup within that population---often a subgroup of racial, gender, cultural, or linguistic minority status, or intersections of these identities. Because ML models are often optimized to achieve high overall accuracy at the population-level, the optimization procedure trains a model that further-marginalizes an already-marginalized group, by down-weighting the relative importance of what little data that group does contribute to the optimization procedure. Thus, the model's \emph{outputs} for the subgroup are bad (i.e., achieve low accuracy, or are of low quality) \emph{because} there is a disparity in the model's \emph{representation} of the subgroup relative to the overall population.
It is significant that Rawls's Difference Principle is operationalized to address this problem of representation disparity because doing so exemplifies a distinction between two types of harms, often posited in \ac{fair ML} literature.\footnote{
    I thank Harini Suresh for encouraging me to discuss this distinction.
}
Recall that I previously noted (briefly) the distinction between \emph{allocative harms} and \emph{representational harms}~\sectionlabel{section:2.2.1}. I now give a fuller introduction to the distinction between allocation and representation.

In \ac{fair ML} literature, allocative harms refer to algorithmic unfairness which \emph{can be} mitigated via resource reallocation, whereas representational harms refer to unfairness which \emph{cannot be} mitigated via resource reallocation~\citep{barocas_fairness_2019}.
Allocative harms are remeidiated by redistributing resources \emph{across} people, whereas representational harms arise from social and political relations \emph{between} people.
Representational harms include stereotyping, misrecognition (e.g., disrespect, indignity, invalidating personhood), denigration (e.g., the use of epithets or other insulting language), and under-representation (e.g., datasets wherein a dominant group is over-represented, typically white men)~\citep{crawford_trouble_2017}.
These harms arise from people's heterogeneous social and political conditions, and so they are difficult to formalize quantitatively  within ML systems.\footnote{
    A commonly given example of representational harm is that of a search-engine which outputs biased results for particular input queries. For example, when searching for ``CEO'' the engine may output only images of white men~\citep{crawford_trouble_2017}, or when searching for ``black girls'' the engine may output content that is demeaning and denigrating to this group~\citep{noble_missed_2012}.
}
These harms are also (in essence) the same as those I described when introducing the sensitivity criterion. In other words, a measure of justice that satisfies the sensitivity criterion would (in theory) also be sensitive to representational harms, which are not remediable by reallocating resources alone.
As I have described it above, the problem of representation disparity seems as though it fits neatly within representational harms (specifically, a harm of under-representation), rather than within allocative harms.
But, as we will see in the following case study, the operationalization of Rawls's Difference Principle takes an allocative approach (i.e., a resourcist approach) to addressing this representational harm, and in doing so it fails to remediate psychosocial injustices that arise from people's heterogeneous social and political conditions.

\subsubsection{The Case of Predicting Sociolects}
\label{section:4.2.1}

Consider ``text auto-complete'' models that predictively suggest the next most-likely word to a user as they are typed in real-time. These models are commonplace on most smartphone text messaging apps, such Apple's iMessage. These models are also known to exhibit linguistic representation disparities for minority sociolects (i.e., socially-restricted dialects). That is, due to disparities in sociolect representation in linguistic training data, text auto-complete models ``allocate'' relatively less predictive text accuracy to minority sociolects, such as so-called African-American English (AAE), while at the same time allocating relatively more predictive text accuracy to majority languages, such as so-called Standard-American English (SAE), thereby reproducing existing discourse inequalities within the ML system~\citep{mayfield_equity_2019}.

In a recent \ac{fair ML} paper presented at the 35th International Conference on Machine Learning (ICML), researchers operationalize one of Rawls's distributive principles to mitigate language representation disparities in a text auto-complete model~\citep{hashimoto_fairness_2018}. In their operationalization, the researchers specify ``predictive text accuracy'' as the ``resource'' for allocation via one of Rawls's distributive rules, what is known as the Difference Principle. The Difference Principle requires that any inequalities in a distribution of resources be to the greatest advantage (maximally beneficial) to the people who are least well-off, measured in terms of their resources~\citep{rawls_theory_1971,wenar_john_2017}. To achieve this, the rule maximizes (over all possible distributions of resources), the minimum (over all people), each person's allocation of resources. Because it ``maximizes the minimum'' in this way, the Difference Principle is also often referred to as the \emph{maximin rule}. Accordingly, when operationalized in a text auto-complete model, the model is optimized to allocate a ``fairer'' quantity of predictive text accuracy across languages and sociolects, which is achieved by allocating less accuracy to the majority language (SAE) and more accuracy to the minority sociolect (AAE), subject to the constraints of the Difference Principle. 

Because the distributive rule accounts for the accuracy disparity between the language (SAE) and the sociolect (AAE), by requiring that any inequalities in the distribution of predictive accuracy be maximally beneficial to the group that is least well-off, such a reallocation of ``resources'' may seem sufficient for achieving algorithmic fairness. But what about the fairness of the measure itself: predictive text accuracy?

Not implausibly, predictive text accuracy could encode biases against particular linguistic groups, depending on the societal contexts in which the text auto-complete model is deployed. For example, it could be that AAE-speakers do indeed prefer increased predictive text accuracy, since text messaging apps often flag AAE words as ``misspelled'' thereby further marginalizing the AAE sociolect in the ``linguistic marketplace'' where dominant languages like SAE enjoy greater linguistic capital and symbolic power~\citep{bourdieu_language_1991}. Alternatively, it could be that AAE-speakers \emph{do not} prefer increased predictive text accuracy for AAE words, since these sociolects are often considered to be ``resistance vernaculars'' that intentionally subvert dominant linguistic norms as a means of celebrating cultural difference and fostering in-group solidarity~\citep{park_appropriating_2008}. In this latter case, highly accurate text auto-complete predictions for AAE words (appearing on, say, a user's iMessage app) may not improve an AAE-speaker's user experience, if those predictions are perceived as culturally appropriating, or if they reproduce psychosocial injustices such as stereotype threat~\citep{steele_stereotype_1995}.\footnote{
 Unfortunately, what AAE-speakers prefer in their text messaging apps remains an open question. Although the researchers took a step toward answering this question by conducting a ``real-world, human evaluation'' of the text auto-complete models, participants in the evaluation consisted entirely of anonymous workers from Amazon's Mechanical Turk, an online labor platform. Demographic information was not collected from these participants, but they were nevertheless tasked with completing the evaluation \emph{as if} they were actual speakers of either AAE or SAE~\citep{hashimoto_fairness_2018}. Because this evaluation did not engage with self-identified AAE-speakers, no conclusions can be drawn about whether they prefer more or less predictive text accuracy in their text messaging apps. 
}

Observe here that predictive text accuracy exhibits the two key characteristics of a resource measure: (1) it is a single-valued quantity that is easy to measure, compare, and rank (by simply looking up the accuracy allocated to each linguistic group), and (2) it is merely a means to the end goal of providing an enjoyable text-messaging user experience. Thus, this ``resource'' measure satisfies the publicity criterion, but fails to satisfy the sensitivity criterion. Although consistently measurable across languages and sociolects, merely reallocating more predictive text accuracy is insensitive to whether AAE-speakers experience that reallocation as helpful and convenient, or stereotyping and insulting. Absent direct engagement with the people whose well-being will be affected by use of this system (in this case AAE-speakers, see footnote 22), it would be a mistake to assume that reallocating more predictive accuracy will necessarily improve, rather than diminish, well-being.\footnote{
 Similar arguments have been advanced against increasing the accuracy of computer vision models used in facial-recognition systems. Similar to representation disparities in linguistic data, computer vision models may be highly inaccurate for minority populations, due to disparities between minority and majority representation in image training data~\citep{buolamwini_gender_2018, bennett_what_2019}. However, the pernicious applications of these models (namely, the increased surveillance and policing of historically and presently oppressed populations) should cast serious doubt on the claim that increasing model accuracy will improve well-being. Whether increasing predictive accuracy is beneficial or detrimental to a person's well-being depends on who that person is and in what context they interact with, or are affected, by the model. Thus, an evaluation of the unfairness or injustice of a particular model will require sensitivity to people's personal heterogeneities societal context.
}

\subsection{``Doing Justice'' to Rawls's Theory}
\label{section:4.3}

The prior cases serve two purposes. First, they substantiate the claim that measures from ML align with resource measures on the two key characteristics (\autoref{tab:measuretable}). Second, they illustrate how measures exhibiting these characteristics can satisfy the publicity criterion, but fail to satisfy the sensitivity criterion, thereby reproducing and entrenching structural and psychosocial injustice. This brings to a close the first part of this paper: the inquiry into what makes the resourcist approach amenable to operationalization in ML systems. And it opens the second part: inquiring into whether the capability approach could be operationalized with more success, and if so, how?

At this juncture, we should pause to ask: what does it mean for a theory to be operationalized successfully? The answer to this question will depend both on the theory selected for operationalization, as well as on how the operationalization is carried out in practice. 
Was Rawls's theory ``done justice'' in its operationalization, so-to-speak? That is, was it operationalized in a way that is true to the original statement of the theory? Following the capability theorists, I have argued that the operationalization of Rawls's theory in the prior cases was unsuccessful. So, I should answer whether that is (A) due to flaws in Rawls's theory, and/or (B) due to how the operationalization was carried out in the particular case.\footnote{
    I thank Sally Haslanger for encouraging me to address these questions.
}

Answering (B), there are at least two ways that the operationalization deviated from Rawls's original theory. First, the \emph{lexical priority} of Rawls's principles requires that the principles of Equal Basic Liberty and Fair Equality of Opportunity are satisfied before moving on to the Difference Principle~\citep{rawls_theory_1971}. (I refer interested readers to a survey by Leif~\cite{wenar_john_2017} for an overview of Rawls's lexically prior principles.) So, if we are true to the original statement of the theory, we will not skip over these principles to pick out the Difference Principle for operationalization by itself. Second, I have previously noted that measures from ML resemble the simple measure of income and wealth, more so than Rawls's primary goods measure, which is broad and inclusive. So again, if we are true to the original statement of the theory, we will also aim to operationalize the primary goods measure and not only the Difference Principle.

Answering (A), given that the operationalization deviated from Rawls's theory in at least two ways, we can ask if an operationalization truer to the original statement of the theory could address the prior cases more successfully. I think a truer operationalization could do better. But, we must also consider the appropriateness of operationalizing Rawls's entire theory of justice in the design of a consumer text messaging app. Would this be ``doing justice'' to Rawls's theory? We can deviate from a theory's statement by failing to operationalize it in its entirety (as above). But, we can also deviate by operationalizing a theory in contexts to which it was not intended to apply (see footnote 3). That is, successful operationalization depends not only on staying true to the theory's statement, but also on applying it where intended, under its intended social conditions. So I won't decide here that Rawls's theory itself is flawed, but only misapplied. These observations may nevertheless count in favor of the capability approach, as we will see later on.

\begin{center}
\huge\textreferencemark
\end{center}

\noindent
Before turning to the capability approach, it is worth addressing another discrepancy between \ac{fair ML} and distributive justice. While discussing the operationalization of Rawls's Difference Principle, I introduced the distinction between harms of \emph{allocation} and harms of \emph{representation}, as it is often posited in \ac{fair ML} literature.
However, I didn't note that philosophers of distributive justice have long-posited an analogous distinction, between injustices of \emph{redistribution} and injustices of \emph{recognition}, where the terms here are used (in essence) in the same sense as allocation and representation, respectively.
In contrast to \ac{fair ML}, the analogous distinction from distributive justice is posited, if not less sharply, then with more nuance.
This is because, for some claims of social injustice (in particular, those relating to disability) it may be difficult to sharply distinguish whether the claim is one of redistribution, or one of recognition.\footnote{
    The sharp distinction between redistribution and recognition becomes uncertain when considering ``reasonable accommodation'' requirements for disabled people under the Americans with Disabilities Act~\citep{united_states_americans_1990}, the The European Union Framework Directive~\citep{whittle_framework_2002}, and/or the United Nations Convention on the Rights of Persons with Disabilities~\citep{united_nations_convention_2008}. Reasonable accommodation requires reconstruction of the public built environment (e.g., offices, restaurants, and parks) to better accommodate disabled people by including ramps, elevators, text in multiple formats, sign-language interpreters, flexible work schedules, and so on~\citep{putnam_disability_2019}.
    In one sense, these accommodations may be characterized as \emph{distributive} in that they require reallocating resources of the aforementioned sort. In another sense, these accommodations partially rectify longstanding \emph{misrecognition} of disabled people as members of society worthy of full participation in public life, with equal access to public spaces. In this latter sense, appealing to an abstract principle of fair or just redistribution may not be required at all. Rather, we may only need to ask what is required for a society to fully recognize disabled people, and reply that it requires (in part) equal access to the public built environment~\citep{putnam_disability_2019}. Thus, recognition requires redistribution, and redistribution aims to secure recognition.
}
Recognition may require redistribution, and redistribution should aim to secure recognition~\citep{putnam_disability_2019}.
The two may be mutually reinforcing, and some philosophers argue that the antithesis between them is a false one.

According to Nancy~\cite{fraser_redistribution_2003}, ``one should roundly reject the construction of redistribution and recognition as mutually exclusive alternatives. The goal should be, rather, to develop an integrated approach that can encompass, and harmonize, both dimensions of social justice.''
Could the same be said of the allocation-representation distinction posited in \ac{fair ML}? In the examples that we have seen, a model's \emph{representation} of a group can produce bad outputs or \emph{allocations} for that group. Here, the distinction between representation and allocation seems to be very narrow, especially when what is being ``allocated'' is something like the model's predictive accuracy.
But, representational harms include more than under-representation in the training data that leads to representation disparity within a ML model. They also include harms that arise from people's \emph{interaction with} that model, after it has been designed and deployed in society, as in the Disability Hiring and Predicting Sociolects case studies. These harms arise from heterogeneous social and political conditions that are not straightforwardly formalizable within the ML model itself, but are nevertheless reproduced and entrenched through the model's misallocation of a particular resource.
Here, allocative harms (i.e., inequities in the distribution of resources) reproduce and entrench representational harms (i.e., those arising from social and political relations between people), and this may occur \emph{even when} resources are distributed equally, as we have now seen many times.

While it is easy to posit allocative and representational harms as two distinct types, it is more difficult to specify the relationship between the two. At the very least, it is clear that they are not mutually exclusive, but mutually reinforcing.
Following Fraser's prescription, we might then ask: how can \ac{fair ML} researchers develop an integrated approach that can ``encompass and harmonize'' both allocation and representation?
Doing so may not be straightforward, or easily operationalized, but we might again look to extant theories of distributive justice for suggestions.
As we have seen, resource measures can be insensitive to heterogeneous social and political conditions from which representational harms arise.
Rawls's primary goods measure attempts to addresses this insensitivity by conceiving of resources broadly and inclusively, but capability theorists argue Rawls's conception still fails to detect and remediate social injustice.
They argue that reallocating resources \emph{alone} is inadequate for addressing structural and psychosocial injustices of recognition (i.e., harms of representation). Doing so requires something \emph{more than} resources.
An approach that integrates allocation with representation may still rely (in part) upon redistributing resources, but it may do so with an aim toward securing broader recognition of people within their society, measured in terms of something \emph{other than} the resources they have in the possession.
Capability theorists argue that recognition is better secured by measuring well-being in terms of what people are \emph{capable} of achieving or not, as members of their society, sensitive to their heterogeneous social and political conditions~\citep{anderson_what_1999, brighouse_justifying_2010}.
\section{The Capability Approach to Measuring Justice} 
\label{section:5}

In his Tanner Lecture lecture, Amartya~\cite{rawls_equality_1979} argued that resource measures are ``concerned with good things rather than with what these good things do to human beings.'' Likewise, the same may be said of the measures discussed in the prior cases. Pre-employment application tests, or ML models with high predictive accuracy, are often assumed to be ``good things'' because, for most people, they are. That is, for most people, it's plausible to assume that these things contribute to people's well-being in the given context (be it searching for job opportunities online or engaging in a text message conversation). 
But, theories of distributive justice aren't only concerned the well-being of \emph{most} people. They're concerned with the well-being of \emph{all} people.\footnote{
 Historically, many widely regarded theories of distributive justice overlooked, intentionally delayed, or disregarded consideration of disability. Up to and including Rawls, no major theory of justice in the western philosophical tradition considered disabled people of central importance~\citep{becker_reciprocity_2005}. After Rawls, nearly all do, in part due to disability-informed critiques of his influential theory. Disability has long been characterized as a ``hard case'' for distributive justice, one that~\cite{rawls_kantian_1975} thought should be addressed eventually, but not centrally, since doing so ``prematurely introduc[es] difficult questions that may take us beyond the theory of justice,'' and ``consideration of these hard cases can distract our moral perception by leading us to think of people distant from us whose fate arouses pity and anxiety.'' Capability theorists disagree. On their view, disability is not ``distant from us'' but in fact widespread~\citep{nussbaum_capabilities_2009}. According to~\cite{rawls_equality_1979}, resource measures such as Rawls's are ``not merely ignoring a few hard cases, but overlooking very widespread and real differences,'' differences which affect the well-being of disabled people first and foremost, but also affect the well-being of all people, since all people ``have very different needs varying with health, longevity, climatic conditions, location, work condition and even body size.'' 
}
Resource measures aim to encompass what contributes to people's well-being, but, capability theorists argue, they miss their mark. Because resource measures are insensitive to personal heterogeneities and societal context, they fail to attend to the different ways some people are (or are not) able to convert resources into well-being. Capabilities theorists argue that what actually contributes to well-being does not therefore lie in the resources themselves, but in what people are \emph{capable} of achieving through their use. The right measure therefore does not lie in the ``good things'' or resources, but in what these ``good things do to human beings''~\citep{rawls_equality_1979}. That is, it lies in people's \emph{capabilities}.

In this section, I first define what capabilities are as they are understood in distributive justice and provide some examples. Then, to distinguish capability measures from resource measures and those from ML, I discuss how they differ on the two key characteristics. Based on these differences, I argue that capability measures satisfy the sensitivity criterion, and they also (in theory) satisfy the publicity criterion. Whether capability measures achieve publicity in practice is somewhat less clear (\autoref{tab:measuretable}).
In upcoming sections, I outline how the capability approach could be operationalized---through a participatory and democratic process---in the design and development of ML systems, to better remediate the social injustices discussed in the previous case studies, and beyond.
While I present the core differences between capabilities and resources, I necessarily omit some subtleties of the capability approach, which has many variants. For a more comprehensive presentation, I refer interested readers to \emph{Wellbeing, Freedom and Social Justice: The Capability Approach Re-Examined} by Ingrid~\cite{robeyns_wellbeing_2017}.

\subsection{What Are Capabilities?}

The capability approach attempts to locate a person's well-being in their capabilities. A person's \textbf{capabilities} are defined as their real opportunities to achieve valuable states of ``being and doing''~\citep{daniels_equality_1990}. States of being might include being sheltered, being nourished, or being cared for. States of doing might include commuting to work, browsing the internet, or communicating with a friend via text message. What does it mean to locate a person's well-being in their capabilities, or their ``real opportunities to achieve valuable states of being and doing''~\citep{robeyns_capability_2016}? Unfamiliar readers will likely find this definition nebulous at first, but capabilities are most readily understood when contrasted with resources.

Recall that because people can use resources to do or to be things that contribute to their well-being, the resourcist approach aims to locate a person's well-being primarily in their resources (i.e., things they can possess or use).
People can usually use their income and wealth to buy a house for shelter, food for nourishment, health insurance for care, a train ticket for their work commute, a computer for internet access, and so on. While capability theorists agree that people can usually use resources to achieve these valuable states of being and doing (being sheltered, being nourished, being cared for, commuting to work, browsing the internet, communicating with friends), they disagree with the implication that resources are therefore the right benefit to look at when measuring what contributes to people's well-being within their society. Why not? Well, according to capability theorists, resources are (as I described earlier) a \emph{means} of achieving those valuable states, but resources are not an end in themselves. That is, although people can usually use, consume, or cash-in their resources to achieve valuable states of being and doing, what actually contributes to well-being are those valuable states themselves, not the means (i.e., resources) used to achieve them~\citep{robeyns_wellbeing_2017}.\footnote{
 In the capability literature, these states of being and doing are called a person's \emph{functionings}. Functionings are the states of being and doing that a person experiences (such as moving freely, being literate, being healthy, being well-nourished, being free from persecution, and so on). Functionings focus ``on the state of the person, distinguishing it from both the commodities [resources] that help generate that state, and from the utilities [welfare] generated by that state''~\citep{sen_capability_1993}. For this reason functionings are sometimes called ``midfare'' because they exist mid-way between resources and welfare~\citep{cohen_equality_1990}.
}
Resources may be a ``proxy'' for a person's well-being, but resources are not equivalent with well-being. As before, being monetarily wealthy is not the same as being well.

We have already seen that, in many cases, the connection between the means (i.e., resources) and the valuable states of being and doing comes apart, depending on people's heterogeneities and societal contexts.
For example, for a person who uses a wheelchair, purchasing a train ticket is not a means to the benefits of public transportation, if the train station lacks wheelchair accessible ramps.
For a blind person, purchasing a computer is not a means to the benefits of internet access, if those technologies are not compatible with screen reader technology.
Similarly, in the prior case studies from ML, a pre-employment application test is not a means to employment if that test is not accessible, and increasing predictive text accuracy is not a means to a better user experience if the predictively-suggested language is stereotyping or disrespectful.
In each of these cases, the resource (train ticket, computer, application test, predictive accuracy) presents a \emph{false} opportunity for achieving the corresponding valuable state of being or doing (commuting by public transit, browsing the internet, being employed, carrying on an enjoyable text message conversation).
That is, what is usually (for most people) an opportunity to convert some resource into a valuable state of being or doing is foreclosed for others, due to their personal heterogeneities and societal context.

Because different people are able to convert their resources into well-being in different ways, and with different degrees of reliability, capability theorists argue that resources alone must not be what contributes to people's well-being---that resources are not \emph{ultimately} valuable to people. Ultimately, they argue, well-being amounts to real opportunities to achieve valuable states of being and doing---and \emph{not} the not-always-reliable resources through which those valuable states may or may not be achieved. The right thing to look at when measuring and remediating social injustice, therefore, is not the quantities of resources people have in their possession, but the states of being and doing that people are capable of achieving, sensitive to their heterogeneous social and political conditions.

\subsection{Two Characteristics of Capability Measures}

Recall the two key characteristics on which resource measures and measures from ML are aligned: (1) these measures can be expressed as single-valued quantities, and (2) what it is that they measure (namely, resources of some sort) are a means to an end, rather than an end in themselves. In this section, I discuss how capability measures differ from these measures on both characteristics, and how they subsequently satisfy the sensitivity criterion and may also (in theory) satisfy the publicity criterion (\autoref{tab:measuretable}).

\subsubsection{Capability Measures Must Be Multi-Valued}

Whereas resource measures \emph{can be} singe-valued quantities, capability measures \emph{must be} multi-valued and can be heterogeneous (i.e., they can combine qualitative and quantitative data). As Martha~\cite{nussbaum_capabilities_2009} puts it, ``it is of the essence of the focus on capabilities to insist that the goods [benefits] to be distributed... are plural and not single, and that they are not commensurable in terms of any single quantitative standard.'' Notice what the essence does \emph{not} insist. It does not insist that the measure is \emph{not} purely-quantitative (i.e., that it is either qualitative or mixed). In other words, a multi-valued quantitative measure is consistent with the capability approach, but so is a purely-qualitative measure or a measure that mixes qualitative and quantitative data. Capability measures can be formulated in many ways (subject to certain process constraints, as we will see later on), so long as they are not single-valued quantities.\footnote{
    However, a formulation that comes close to a single-valued quantity might also contravene the essence of the approach. For example, a double-valued quantitative measure, while technically permissible because it is not commensurable in terms of any single quantity, may not be capture the fullness and potential of the approach. This is because each value in the measure is an \emph{indicator} for a capability that people have in a given context. There will be few values in the capability measure only when there are few capabilities relevant to that context. In most contexts of practical concern, however, there will be many relevant capabilities.
    For example, consider the context of remediating social injustice relating to ambulatory disability. A single-valued resource measure might aim to measure whether people with ambulatory disabilities have in their possession a particular resource (or bundle of resources), including income and wealth, wheelchairs or prostheses, and so on. By contrast, a multi-valued capability measure might aim to measure whether people achieve the capability of ``freely moving without barriers'' within their society, regardless of disability. Achievement of this broader end might be measured through multiple indicator values, such as whether public spaces are wheelchair accessible, whether public transportation is widely available, and whether people can acquire the relevant resources, including income and wealth, wheelchairs or prostheses, and so on.
}

Here, it is important to state explicitly: the capability approach does not exclude resources from playing a constituent role in the formulation of a capability measure. Exactly the opposite: the approach (in part) \emph{relies on} reallocating resources so that people achieve greater capabilities.
Why?
Capabilities are based on people's states of being and doing. Because these states of being and doing are coincident with people, they cannot be allocated in any direct fashion, only measured~\citep{brighouse_justifying_2010}.
Thus, the difference between the resourcist and capability approach is (in part) one of emphasis. The former emphasises the resources people have in their possession, and usually measures them through single-valued quantities. The latter emphasises what it is that people are capable of doing with those resources given their heterogeneities and societal context, and measures this through multiple values.
In both cases, resources are reallocated.
This begins to suggest how capability measures might be employed to develop an integrated approach that can ``encompass and harmonize'' allocation and representation (i.e., redistribution and recognition), as prescribed earlier by Nancy~\cite{fraser_redistribution_2003}.

\subsubsection{Capabilities Are Ends, Not Means}
Recall that in the resources-versus-capabilities debate, an important distinction is posited between things that are a ``means to an end'' and things that are an ``end in themselves.'' In this way, it is argued that resources  are a means of achieving well-being, but resources are not \emph{ultimately} valuable to people, as an end in themselves.
We have now seen a number of examples in which resources fail to confer their intended benefit, due to context-sensitive variations in how reliably people convert resources into well-being.
Because resources are a \emph{means} to achieving well-being, acquiring more of them may not lead to corresponding improvements to people's quality of life and future prospects.
In these cases, there is a ``slippage'' between the resources a person has in their possession and how those resources affect their well-being.

Resources may be an easily-measurable ``proxy'' for well-being, but they may be unreliably beneficial.
Thus, resource measures may fail to encompass well-being because they do not seek to measure well-being \emph{itself} (i.e., what is ultimately valuable to people)~\citep{robeyns_capability_2016}.
By contrast, the capability approach offers an alternative framework for conceptualizing well-being itself, one that understands well-being in terms of a person's valuable states of being and doing~\citep{robeyns_capability_2003}.
These states of being and doing are ultimately valuable to people. That is, there is no way in which a people's capabilities are ``converted'' into anything of \emph{greater} value.
Their value is final.
People's capabilities are thus an end in themselves, not a means.
Because of this, there is no ``slippage'' between means and ends in the capability framework of well-being.
To measure well-being, capability theorists aim to measure people's capabilities, sensitive to their heterogeneities and societal context.

\subsection{Capability Measures: Sensitivity, And Publicity?}

Capability measures differ from resource measures and measures from ML on their two key characteristics: (1) they must be multi-valued, and (2) what is being measured (namely, capabilities) are an end in themselves, not a means. Taken together, these characteristic differences contribute to a measure that satisfies the sensitivity criterion, and may also (in theory) satisfy the publicity criterion.

Consider sensitivity. The means-end distinction relates to sensitivity as follows. 
Because capabilities are ends in themselves and measured with respect to people's heterogeneities and societal context, they are sensitive to the heterogeneous social and political conditions under which people live.
Additionally, the multi-valuedness of a capability measure contributes to its sensitivity as follows.
Because capability measures are necessarily multi-valued, they achieve greater sensitivity to heterogeneous social and political conditions via multiple indicators.
By contrast, resource measures are often expressed as single-valued quantities and therefore offer a simple---perhaps overly-simplifying---measure of well-being.
However, including multiple indicator values is not enough to provide a fuller measure of well-being.
It also matters \emph{which} values are selected for inclusion in the capability measure.
Here, the capability approach prescribes selecting values that are indicators of people's capability achievement within their society, rather than adopting a general-purpose resource standard. 

Consider publicity. Because capability measures must be multi-valued (and are possibly heterogeneous), they will not permit strictly unambiguous comparisons and rankings (i.e., a complete interval rank, not only a partial ordinal ranking).
While it is clear that a capability measure will not provide a public standard of measurement that is \emph{equally} as strict as that provided by a singular quantitative resource measure, it does not necessarily follow that a capability measure \emph{cannot} be publicly legible and verifiable.
This will depend on the requirements set for public legibility and verifiability. A capability measure \emph{can} provide a standard of measurement of some sort. The question is whether that standard will be strict enough for the purposes of remediating social injustices. 
Resourcists argue that the capability standard of measurement is not strict enough; that justice demands interpersonal and intercontextual comparisons be clear, and not murky~\citep{brighouse_equal_2010,brighouse_critique_2010}. 
Capability theorists counter-argue, in two ways.

First, capability theorists observe that there may be a trade-off between improving human capabilities and achieving a publicly legible and verifiable standard of measurement. Achieving full publicity might require omitting important contextual information, and doing so could allow social injustices to slip by undetected and unremediated, as we have seen.
Conversely, admitting too much contextual information into the measure might strain public legibility, and doing so could also perpetuate social injustices by stymieing oversight and accountability of the systems and institutions regulated by that measure.
When the trade-off is between improving human capabilities and achieving a fully legible and verifiable measure, capability theorists argue that we should opt for improving human capabilities, by relaxing the strictness of the publicity criterion~\citep{brighouse_two_2010}.
In other words, so long as capabilities are more-or-less equalized across people in society, a capability measure does not need to achieve publicity fully, but only partially.

Second, capability theorists appeal to the way in which the capability measure is specified: through political consensus; namely, a participatory and democratic process. When a measure of justice is specified through such a process, it achieves public legibility and verifiability of a different sort. In the next section, I will elaborate further on what is meant by a process that is participatory and democratic.
\section{Operationalizing The Capability Approach} 
\label{section:6}

I now address the question of whether the capability approach can be operationalized in the design and development of ML systems, and if so, how? The answer to this question will depend (in part) on how strictly the requirement for public legibility and verifiability is drawn. Capability measures can (in theory) satisfy both sensitivity and publicity criteria. Sensitivity is achieved through a multi-valued (possibly heterogeneous) measure of people's capabilities. Publicity is achieved by either relaxing the strictness of the requirement for public legibility and verifiability (in favor of improving human capabilities), and/or by specifying the capability measure through a participatory and democratic process. 

Before elaborating on what this means, I first note that such a process is not impracticable. For decades, the capability approach has been operationalized as an alternative to standard single-valued measures of well-being in welfare economics (e.g., measures of Gross Domestic Product and Gross National Product). It has been operationalized in the United Nations Human Development Index, to measure international capabilities in health, education, and income~\citep{fukuda-parr_human_2003, fukuda-parr_handbook_2009}. And, of particular relevance to this inquiry, the approach has been operationalized in the design of many information and communications technologies~\citep{oosterlaken_evaluating_2012}. For an in-depth survey of technology-specific applications, I refer interested readers to the anthology \emph{The Capability Approach, Technology, and Design} by Ilse Oosterlaken and Jeroen van der Hoven~\citep{oosterlaken_capability_2012}.

\subsection{How to Do It: A Five-Step Process}
\label{section:6.1}

What does it mean to specify a capability measure through a participatory and democratic process?
And how would the resulting measure more successfully remediate the social injustices from the prior case studies, and beyond?
I emphasize that there are no \emph{general} answers to these questions. Different capability theorists propose different methods of operationalizing the approach~\citep{robeyns_sens_2003}. As just one example of how such operationalization might proceed in a participatory and democratic fashion, I outline a five-step process, extrapolated from Murphy and Gardoni~\citep{oosterlaken_design_2012}.

\begin{quote}
\textbf{Step 1.} In collaboration with the affected parties (i.e., people whose well-being may be affected by the system under consideration), select the set of relevant capabilities (i.e., what encompasses people's well-being with respect to that system).
\end{quote}

\begin{quote}
\textbf{Step 2.} For each capability in the set, select indicators for that capability. Indicators might include single-valued quantities, open-ended written testimony, and/or likert scale ratings, indicating the extent to which its corresponding capability is achieved.
\end{quote}

\begin{quote}
\textbf{Step 3.} Convert each indicator into an index that ranges from 0 (minimum achievement) to 1 (maximum achievement). Note: this step is optional, if it is preferred that the data remain qualitative or mixed.
\end{quote}

\begin{quote}
\textbf{Step 4.} Combine all indices to create an aggregate measure (e.g., by averaging, if the measure is quantitative).
\end{quote}

\begin{quote}
\textbf{Step 5.} Iteratively design and evaluate the system to bring all affected parties to a \emph{baseline threshold} level of capability achievement. If and when all parties reach the baseline threshold, increase the threshold.
\end{quote}

\noindent
In this way, the capability approach specifies the distributive principle of ``equal capabilities'' across affected parties, with respect to the particular system and context under consideration. Note that this five-step participatory process could result in an aggregate, purely-quantitative measure, but it need not. For example, converting open-ended written testimony into a single-valued quantity at Step 3 is not required, as it may be difficult to make this conversion in any meaningful way, or it may be more valuable and illuminating to keep qualitative data in its qualitative form.\footnote{
 Importantly, even if this process did result in an aggregated quantity, this multi-valued quantitative measure would not be reducible to, or give priority to, any constituent single-valued quantity. This is because, in Step 2, if a single-valued quantity were selected as an indicator of some relevant capability, this indicator would be only one of many equally important indicators, each measuring a corresponding relevant capability, of which there are necessarily more than one. Thus, the single-valued quantity would not be the primary measure of whether or not the system is fair, just, improves or diminishes the well-being of those who are affected by it. Additionally, in Step 4, the aggregation of multiple single-valued quantities should be done in a way that strongly prioritizes any particular quantity. The most basic approach to aggregation is un-weighted averaging.
} The process only stipulates that the capability measure is multi-valued and not singular. 

Consider again the prior cases from ML. Operationalizing the capability approach in the case of Disability Hiring would first require specifying the set of relevant capabilities in collaboration with the users of online hiring platforms, requiring representation from users who are likely to use or interact with those systems. This would include people who navigate the hiring platforms using screen reader technology. 
Here, it is important to note that participatory democracy does not imply majoritarian rule, or even rule by plurality.
That is, the set of relevant capabilities should be determined through a participatory and democratic process, sensitive to both majority and minority interests.
(No doubt, such a process introduces a number of difficulties and limitations, the discussion of which I will postpone until the next section.)
Hypothetically, the relevant capabilities selected through such a process might include: successfully completing an employment application test (as indicated by whether or not a user files the test without reporting difficulties), having the option to communicate directly with a human hiring agent (as indicated by the presence or absence of customer support on the hiring platform), and so on.
Similarly, operationalizing the capability approach in the case of Predicting Sociolects would first require specifying the set of capabilities relevant to text auto-complete models, in collaboration with the users of those models---particularly, self-identified speakers of AAE---prior to optimizing the model to achieve a particular accuracy benchmark.
Hypothetically, the process-determined set of relevant capabilities might include: having an enjoyable user experience (as indicated by user testimony or likert scale ratings), contesting or customizing the predictions of the text auto-complete model (as indicated by whether these controls are exposed in the user interface), and so on.

The aforementioned capabilities and indicators primarily exist at the interface between people and the ML system, and hypothetically result in a measure that is multi-valued and heterogeneous (i.e., combining quantitative and qualitative data).
In this sense, the capability approach is analogous to research methodologies often employed in the field of human-computer interaction (including ethnography, participatory action research, and other mixed methods), whereas the resourcist approach is analogous to methodologies often employed in ML.
But here it may be wondered: when operationalized in the design and development of ML systems,  would the capability approach apply at ($I$) the level of algorithm design and model optimization (as in much of current \ac{fair ML} research), at ($II$) the level of human interaction with the ML system (as in the examples above), or at ($III$) the level of societal well-being (as in traditional welfare economics)?
The answer is, potentially, any or all of the above, depending on the scale and scope of the ML system under consideration: the relevant capabilities and the evaluation of justice or fairness should be scoped to address that particular system.
That is, when operationalized with respect to a particular ML system, the approach should correct for capability deficits resulting from (relating to) that system, and not necessarily beyond.\footnote{
I thank Reuben Binns for encouraging me to clarify this point.
}

When designing algorithms at level $I$, it may be unrealistic or impracticable for ML developers to correct for capability deficits at level $III$, where the set of relevant capabilities may be different because these deficits do not arise from ML systems.
However, considerations of societal well-being may still factor into participants' selection of relevant capabilities for a particular ML system.
For example, level $III$ considerations might lead participants to conclude that a particular ML system (scoped to level $I$ and/or $II$), should not be built or deployed at all, if that system seems likely to exacerbate capability deficits at the level of societal well-being.
Thus, while the approach technically permits a capability measure that is purely-quantitative---applied at level $I$ alone---such a measure would leave behind part of what makes capability measures useful to begin with: they can (in theory) bridge these different levels of evaluation, from ($I$) algorithm design, to ($II$) human-system interaction, to ($III$) societal well-being.
And they can do so in a principled fashion, under a unified conceptual framework of well-being.
Here again, this suggests how capability measures can be employed to develop an integrated approach that can ``encompass and harmonize'' allocation and representation (i.e., redistribution and recognition), as prescribed earlier by Nancy~\cite{fraser_redistribution_2003}.
In the next section, I discuss some of the most noteworthy advantages and limitations of the capability approach.

\subsection{Advantages and Limitations}
\label{section:6.2}

The five-step participatory process outlined in the previous section raises a number of questions. Who are the affected parties? How do these people select the relevant capabilities, the indicators for those capabilities, and the baseline threshold of capability achievement? How is data collected and aggregated?  One short, simplifying answer is that each of these decisions is to be made through a participatory and democratic process~\citep{sen_quality_1993}. The longer, more complicated answer is that such processes are hardly straightforward, and each decision point poses further challenges to adopting the capability approach in practice. Here, I will not attempt to address all of these questions; the capability literature is vast and there are many answers to these questions on offer~\citep{robeyns_wellbeing_2017}. Instead, I discuss some of the advantages and limitations of the approach as they relate specifically to the design and development of ML systems.

\subsubsection{Sensitivity Across the Data Pipeline}
\label{section:6.2.1}

There are at least two advantages to operationalizing the capability approach in ML systems. First, the approach is underspecified by design. At minimum, it stipulates no more than a \emph{conceptual framework} for developing capability measures of well-being in a context-sensitive fashion~\citep{robeyns_wellbeing_2017}. Contrast this with Rawls's theory of justice, which is elaborate, highly-specified, and intended to apply under certain ideal social conditions, not directly operationalized in practice (see footnote 3). The capability approach is \emph{intended} to be operationalized in a variety of real-world contexts, and is underspecified in order to be flexibly applied. As we will see, underspecificity has its limitations, but for now let's consider its advantages.

Because it is underspecified, the capability approach has been flexibly operationalized for a variety of projects, contexts, and heterogeneous data types, both qualitative and/or qualitative~\citep{alkire_valuing_2005}. Applications of the approach have ranged from a quantitative evaluation of gender differences in India~\citep{sen_commodities_1985}, to a qualitative ``asset mapping'' survey seeking to understand and improve the quality of life for people in low-income neighborhoods in California~\citep{jasek-rysdahl_applying_2001}, to the  aforementioned technology-specific applications.
This breadth of applications suggests that the approach may be suited to answering recent calls for incorporating qualitative and heterogeneous data into the evaluation of ML systems~\citep{malik_hierarchy_2020}.
So far, ML systems have been largely evaluated using standard quantitative measures on benchmark datasets. And, although often critiqued as insensitive to societal context~\citep{selbst_fairness_2019}, it has been largely unclear \emph{how} to achieve greater sensitivity to people's heterogeneous social and political conditions, in a principled fashion.
The capability approach provides a conceptual framework that can (in theory) bridge several levels of evaluation: the algorithmic, the human-system interface, and the societal.

A second advantage of the capability approach is that it is process-oriented. That is, it does not at the outset aim to achieve any predetermined outcome, such as optimizing a particular measure, maximizing the accuracy of an ML model, and so on.
Instead, the approach remains open to any and all outcomes, sensitive to the interests, needs, personal heterogeneities, and societal contexts of participants who engage in the process of developing the capability measure.
In this way, when operationalized in the design of information and communications technologies, the capability approach can (in theory) achieve sensitivity from the beginning, to the end, of the data pipeline.
That is, not only does a measure achieve sensitivity to people's heterogeneities with respect to particular extant datasets~\citep{gebru_datasheets_2020}, it aims to achieve sensitivity across the entire data pipeline: \emph{who} collects the data, \emph{what} data is collected, \emph{how} data is being collected, and \emph{which} calculations and evaluations are made using the data.
In this sense the capability approach is reminiscent of value-sensitive, participatory, or inclusive design practices~\citep{frediani_processes_2012, oosterlaken_human_2015}, as well as more recent calls for justice-based and feminist design practice in data science~\citep{costanza-chock_design_2018,costanza-chock_design_2020,dignazio_data_2020}. On process-oriented variants of the approach, the foregoing questions---who, what, how, and which---are to be answered through a participatory and democratic process, such as the one outlined above.

\subsubsection{Selecting the Relevant Capabilities}
\label{section:6.2.2}

While the underspecified and process-oriented design of the capability approach has its advantages, it also brings with it many limitations.
In particular, the question of how to select the relevant capabilities is a prominent point of tension among capability theorists, and opens the approach to criticism from resourcists concerning the public legibility and verifiability of a capability measure.
There are many answers to this question on offer, the most prominent of which I discuss in this section.

Sen has repeatedly declined to endorse a specific set of relevant capabilities, preferring instead that the approach remain an open-ended conceptual framework, and not a well-defined theory.
According to Sen, the relevant capabilities should be decided via democratic participation and political consensus, sensitive to the heterogeneous social and political conditions of the people who will be affected by the resulting measure~\citep{sen_capability_1993}.
In practice, however, achieving political consensus may be slow, difficult, or expensive.  Questions of scope and scale are likely to arise. Participants may disagree on which capabilities are relevant, and their valuations of particular capabilities may be biased by their own \emph{adaptive preferences}, whereby they unconsciously downgrade the importance of capabilities previously inaccessible to them~\citep{alkire_concepts_2008, khader_adaptive_2011}.\footnote{
    I thank Sally Haslanger for suggesting these references.
}
To address some of the practical difficulties with participatory and democratic process, Robeyns has proposed pragmatic criteria for guiding the selection of relevant capabilities in policy and empirical research, modulated by the scope and scale of the system under consideration~\citep{robeyns_capability_2003}. 

In contrast with Sen's view, Nussbaum and Anderson have both proposed specific and highly-abstract capability sets relevant to their particular areas of concern. Anderson proposes that the relevant capabilities are whichever allow a people to function as equal citizens in a democratic society~\citep{brighouse_justifying_2010}.
Nussbaum proposes a general set of \emph{ten core capabilities}, which can be translated into more detailed and specific capability sets depending on local context~\citep{nussbaum_women_2000}.\footnote{
 Nussbaum's set of core capabilities includes \emph{bodily health} (i.e., nourishment and shelter), \emph{bodily integrity} (i.e., freedom of movement, freedom from violence), \emph{emotions} (i.e, being able to have attachments to things and people), \emph{practical reason} (i.e, being able to engage in critical reflection about the planning of one's life), \emph{affiliation} (i.e, being able to live with others), \emph{play} (i.e., being able to laugh and pursue recreational activities), and \emph{control} over one's environment (i.e, political choice and participation)~\citep{nussbaum_frontiers_2009}, among others. Only some of these capabilities may be relevant to the design and development of ML systems.
}
While these capability sets are proposed for their particular application areas, and so are not immediately applicable to the design and development of ML systems, they may nevertheless guide the identification of capabilities relevant to these systems.
For example, automating important decisions (such as those regarding employment, loans, or healthcare) in ways that are opaque may significantly inhibit a person's capability for \emph{practical reason} (one of Nussbaum's core capabilities). That is, doing so may inhibit a person's ability to critically reflect and plan their life~\citep{coeckelbergh_health_2010}. If a person does not know for what reason they were denied employment or a loan, it will likely be difficult for them to self-reflect and better plan for the future job or loan applications.
Similarly, automating these decisions in ways that do not provide opportunities to contest their outcomes may inhibit a person's capability for \emph{control} over their environment (another one of Nussbaum's core capabilities).

The underspecified design of the capability approach leaves it open to criticisms of impracticability and public illegibility, when the participatory process fails to result in an agreed-upon capability set. This may be so, but public deliberation and democratic process is never guaranteed at the outset to arrive at definitive answers. If answers could be arrived at, then we (i.e., participants in the process of specifying the capability measure) could have some confidence that the resulting measure would be sensitive to our heterogeneous social and political conditions. Alternatively, if no such agreed-upon answers are found, engaging in a participatory and democratic process may---at the very least---provide some valuable insight into the question of which capabilities are relevant to the design and development of ML systems.

\subsection{Capability Measures in Machine Learning}
\label{section:6.3}

Operationalizing the capability approach in the design and development of ML systems can (in theory) satisfy the sensitivity and publicity criteria for a measure of justice in machine learning. Sensitivity is achieved through a multi-valued (possibly heterogeneous) measure of human capability with respect to these systems.
Publicity is achieved by either relaxing the strictness of the requirement for public legibility and verifiability (in favor of improving human capabilities) and/or by specifying the measure through a participatory and democratic process. Thus, whether or not the capability approach can be successfully operationalized in the design and development of ML systems will depend (in part) on how strictly the requirement for public legibility and verifiability is drawn, and on the particular details of the process.

If the field of \ac{fair ML} draws the requirement for public legibility and verifiability strictly (i.e., by requiring an rank interval ordering of well-being, rather than a partial ordering), then a capability measure may not be adequate for remediating injustice in machine learning.
However, I have noted that the capability approach has been successfully operationalized in a variety of fields (some of which are not less quantitatively-strict than ML), and the approach has at least been given passing consideration by some \ac{fair ML} researchers~\citep{gajane_formalizing_2017, jurgens_just_2019}.
Additionally, recent, complementary work has proposed how to better contextualize ML models and datasets with respect to their societal context~\citep{mitchell_model_2019, gebru_datasheets_2020}, how to engage in participatory approaches when designing and developing ML systems~\citep{kulynych_participatory_2020}, and how a different but related conception of ``capabilities'' can be operationalized in the evaluation of ML systems~\citep{ribeiro_beyond_2020}.
These examples suggest that, if the publicity requirement is relaxed within reason, a capability measure may better encompass what it is that contributes to people's well-being, as it relates to these systems, and better detect and remediate the social injustices that arise from their deployment.

Toward that end, I have outlined one way the capability approach could be operationalized in the design and development of ML systems, through a five-step participatory and democratic process. I have also discussed the most noteworthy advantages and limitations of such a process. While its success cannot be guaranteed at the outset, I have suggested that even if failing to satisfy strict criteria for public legibility and verifiability, engaging in a participatory and democratic process may still yield information relevant and valuable to the urgent remediation of social injustices arising from the deployment of ML systems. It may, therefore, still be worth the effort.

\section{Conclusion}
\label{section:7}

In this paper, I have extended---from philosophy to machine learning---a longstanding debate between defenders of the resourcist and capability approaches to measuring justice. By establishing a theoretical \emph{resemblance} between resource measures and measures from \ac{fair ML}, where algorithmic unfairness has often been conceptualized as a problem of resource allocation, I have argued that the capability theorists' critiques of the resourcist approach carry over to ML systems.
In two real-world case studies, I have shown how the measures selected for mitigating algorithmic unfairness can end up entrenching and reproducing social injustice---just as capability theorists argue resource measures do. 
I have aimed to provide a constructive critique of (and partial corrective to) the significant attention \ac{fair ML} researchers have already paid to the resourcist approach (Rawls's theory, in particular).
This critique is constructive because I do not merely criticize the operationalization of Rawls's theory in ML systems. I also introduce an alternative approach, perhaps better suited to such operationalization.
I have qualified this critique by noting where and how operationalized analogues of Rawls's theory have diverged from its original statement and intended application. 
I have outlined what operationalizing the capability approach in ML systems could look like in practice---through a participatory and democratic process---and discussed how doing so could better remediate the social injustices from the case studies, and beyond. 
Finally, I have discussed some of the most noteworthy advantages and limitations of the capability approach.

Extending the longstanding resources-versus-capabilities debate from philosophy to ML has revealed some broadly applicable characteristics of (and criteria for) a measure of justice in machine learning.
\begin{itemize}
    \setlength\itemsep{0.0em}
    \item Whether a measure is single or multi-valued; quantitative and/or qualitative.
    \item Whether what is being measured is a ``means to an end'' or an ``end in itself.''
    \item Whether or not a measure is sensitive to people's heterogeneities and societal contexts.
    \item Whether or not a measure is publicly legible and verifiable.
\end{itemize}
\noindent
These characteristics and criteria have broad implications for the practicability of quantitative and/or qualitative evaluations of justice and fairness in machine learning, for the participatory design of these systems, and for the democratic oversight over the institutions that build and deploy them.

\subsection{Replying to Objections}

The foregoing inquiry remains open to some objections, to which I reply in this final section. Specifically, the reader may (1) object to the established resemblance between resource measures and those from ML, (2) object to the capability theorists' critique of resource measures, (3) object to the capability approach from more conservative perspectives, and/or (4) object to the approach from more critical perspectives. I will reply to these objections in reverse order.

Regarding (4), the capability approach has been criticized from more critical perspectives as inadequately attentive to how social and political institutions produce and reproduce power~\citep{robeyns_capability_2003}. The concern here is that, although the approach engages a participatory and democratic process, it does not necessarily attend to the ways that participation in that process may be coerced by institutional power, or biased by participants' adapative preferences~\citep{alkire_concepts_2008, khader_adaptive_2011}. Without attention to how institutions exert power over participants, participatory and democratic processes may reduce to perfunctory representation, ``tokenism'', or ``participatory theater'' that does not substantively address and improve participants' well-being. Robeyns has responded to this concern by appealing the underspecified design of the approach~\citep{robeyns_capability_2003}.  Because it stipulates at minimum no more than a conceptual framework in which capabilities are taken as the right measure of justice, a variety of more critical and progressive theories may be incorporated into this framework. Specifically, in this case, theories that address power differentials in democratic deliberation may be incorporated, as needed.

Regarding (3), the capability approach has been criticized from more conservative perspectives, as being impracticable on the basis of parsimony. Operationalizing the approach is likely to be expensive, perhaps so much so that profit-motivated, fiscally conservative institutions and corporations will decline to adopt it in practice. Critiques from parsimony may have some merit for small, funding-scarce projects. But, I argue (briefly) they do not apply to ML, for two reasons. First, ML is a highly-funded, highly-profitable area of industry and academic research. The expense can be spared. Second, the societal benefits of building and deploying ML systems is not only unproven, but highly dubious. Their detriments are more readily apparent. Showing that deploying these systems improve well-being and remediate---rather than reproduce---social injustices will require careful, comprehensive evaluation. Arguably, any adequate evaluation (capability-based or otherwise) will be time-consuming and monetarily expensive.

Regarding (2), in the resources-versus-capabilities debate, some philosophers argue that the two measures are ultimately not all that different~\citep{brighouse_equal_2010}. Or, more commonly, they argue that that the resourcist approach can address the capability theorists' critiques in various ways.
While resource measures are not sensitive in the same way as capability measures, resourcists have argued that this insensitivity can be addressed through other means, for example through social welfare programs that provide compensatory reallocations of resources on the basis of disability, unemployment, age, and so on~\citep{brighouse_critique_2010}.
Additionally, when considering the totality of Rawls's theory, some argue that a measure of justice doesn't need to be sensitive to heterogeneous social and political conditions because Rawls's lexically prior principles (i.e., equal basic rights and liberties, fair equality of opportunity) and his measure of primary goods, will address the cases on concern to the capability theorists~\citep{rawls_political_1993}.
Related to this last point, it remains an open possibility that, if operationalized in its totality---rather than piecemeal---in ML systems, the resourcist approach could better address the prior cases from ML. Although, it still remains unclear whether such an operationalization would be appropriate to those particular contexts~\sectionlabel{section:4.3}.

Finally, regarding (1), the objection that I have not sufficiently established the resemblance between resource and measures from  ML, I do think that a fuller characterization of the latter is possible. Of the prominent approaches from theories of distributive justice (resourcist, capability, welfarist), I maintain that the measures from ML most-closely resemble those belonging to the resourcist approach, to the extent that they align on the characteristics and criteria previously discussed. However, this resemblance does not preclude the possibility of a different type of measure or approach, distinct from these but yet-to-be specified. Inquiring into this possibility is not the primary aim of this paper, and is left to future work.
\section{Acknowledgements}

For supporting, supervising, and evaluating this paper, I am especially grateful to Arvind Satyanarayan, Sally Haslanger, and Reuben Binns. 
For providing detailed feedback and edits, I am very grateful to Marion Boulicault.  For providing helpful comments and suggestions, I thank Harini Suresh, Milo Phillips-Brown, Gabriel Karger, and Jason White. For reading early drafts, I thank Aspen Hopkins, Crystal Lee, and Jonathan Zong. I thank the reviewers and participants from the 2020 ACM Conference on Fairness, Accountability, and Transparency, and from the 2019 ACM SIGACCESS Conference on Computers and Accessibility: Workshop on AI Fairness for People with Disabilities. I thank Momin M. Malik whose template I borrowed for typesetting this document. This paper is based upon work supported by the National Science Foundation Graduate Research Fellowship under Grant No. 1122374.

\bibliography{references.bib}

\end{document}